\documentclass[%
aip, POP,
 sd,%
 amsmath,amssymb,
preprint,%
]{revtex4-1}

\usepackage{lineno,hyperref}
\usepackage{bm}% bold math
\usepackage{graphicx}% Include figure files
\usepackage{subcaption}
\modulolinenumbers[5]

\usepackage{xcolor}

\begin{document}

\title{Effects of collision enhanced charging on dust crystal}

\author{Althea Wilson}
\email{agw0001@uah.edu.}
\author{Babak Shotorban}%
 \email{babak.shotorban@uah.edu.}
\affiliation{ 
Department of Mechanical and Aerospace Engineering, 
The University of Alabama in Huntsville, Huntsville, AL 35899
%\\This line break forced with \textbackslash\textbackslash
}%

\date{\today}% It is always \today, today,
             %  but any date may be explicitly specified
\begin{abstract}
Numerical simulations of monolayer dust crystals in an RF complex plasma were performed to examine the crystal structure and quantify the effects of including the collision enhanced ion current in the charging model. A GEC cell similar to a previous experimental work was modeled for a range of RF voltages, using a continuum description for the plasma and a particle description for dust grains. The time history of each dust grain was monitored. The dust charge was computed using both the OML and the collision enhanced charging (CEC) model applicable to the sheath region. The dust model accounted for the electric force, ion drag force, neutral drag force, gravity, and the ion wake. The CEC model produced a lower charge and lower electric force which agreed better with the experimental data. Then dust crystals composed of 40 to 100 grains were modeled and the levitation height and inter-particle spacing of the resulting crystals was examined. Including the collision enhanced current reduced the inter-particle spacing but only had a minor effect on the levitation height. 

This article may be downloaded for personal use only. Any other use requires prior permission of the author and AIP Publishing. This article appeared in A.~Wilson and B.~Shotorban, Physics of Plasmas {\bf 28}, 033705 (2021) and may be found at \url{https://doi.org/10.1063/5.0031596}
\end{abstract}

%\pacs{Valid PACS appear here}% PACS, the Physics and Astronomy
%                             % Classification Scheme.
%\keywords{Suggested keywords}%Use showkeys class option if keyword
%                              %display desired
\maketitle

\section{\label{sec:Intro}Introduction}  %%% Introduction

Low temperature plasmas containing micro- or nanometer-size solid particles, are common in both nature and industrial setups \cite{Shukla2001}. Such complex (or dusty) plasmas
naturally occur in many astrophysical phenomena such as nebula, comets and planetary rings \cite{Goertz1989}. Protoplanetary disks are dusty plasmas where the aggregate-plasma interaction has significant effects on the formation and properties of the dust aggregates \cite{Matthews2011}. 
Dusty low temperature plasmas are also common in manufacturing setups where plasmas are utilized to fabricate semiconductors or coat micro-particles.  
Further, microelectronics manufacturing is concerned about contamination due to dust \cite{Selwyn1990}, which is the byproduct of laser etching or chemical reactions and can degrade the semiconductors if the grains settle on the chip.  
On the other hand, in materials processing applications such as plasma enhanced chemical vapor deposition (PECVD), dust grains in the form of micro- or nano-particles are deliberately introduced to be coated \cite{Cao2002}. 
PECVD is also used to deposit films onto substrates \cite{Stoffel1996} and grow carbon nanotubes \cite{Meyyappan2003}.

The observation of Coulomb crystallization \cite{Ikezi1986,Thomas1994} in dusty plasmas inspired additional interest. In Coulomb crystallization, the dust grains form an organized structure that is suspended in the plasma. 
In RF plasmas, small dust crystals formed under the influence of gravity are flat \cite{Thomas2005} and only a few layers thick while crystals formed in micro-gravity are three-dimensional and often contain central voids \cite{Goedheer2008,Schwabe2014}.  The horizontal structure is typically hexagonal, however the vertical structure is aligned rather than close-packed \cite{Schweigert1996}. The crystal structure interests physicists because it allows solid state physics to be examined on a scale accessible with optical techniques as well as displaying some unique features such as melting \cite{Schweigert1998,Rubin2006}.
The phase transition between crystalline structures and liquids is related to the energy of the grains as well as the presence of structural defects~\cite{Klumov2010} and is an area of ongoing research for both two-dimensional~\cite{Hartmann2010,Klumov2019} and three-dimensional~\cite{Steinmuller2018,Reshetniak2019} systems.
Experimental observations of Coulomb crystals have examined dislocations \cite{Nosenko2011}, density waves \cite{Schwabe2007} and the dust acoustic wave \cite{Barkan1995,Merlino1998} among other topics.   

In the modeling of dusty plasmas,  the dust is either treated as a continuum medium (fluid description) \cite{Goedheer2009} or  dust grains are tracked in the Lagrangian framework (particle description). 
A fluid description of the dust assumes the presence of a large quantity of dust so it does not typically deal with relatively small crystals or provide information on the crystal structure.
On the other hand, the spatial distribution of the dust density distribution is readily calculated in the fluid description so it is useful for investigation of phenomena such as void formation in microgravity experiments \cite{Land2010} where a precise description of the crystal structure is usually not of interest. 

A critical aspect of modeling dusty plasmas is the charging representation.
The traditional method of determining the ion and electron currents is orbital motion limited (OML) theory \cite{Allen1992} which assumes a collisionless plasma where the electrons have a Maxwellian distribution and the ions are treated assuming either a Maxwellian distribution, for the bulk of the plasma, or a monoenergetic beam, for the sheath\cite{Barnes1992}. Sometimes, the OML model is modified to use a shifted Maxwellian distribution for the ions \cite{Douglass2011} which allows the model to handle both the bulk and sheath. However, the OML model neglects charge exchange collisions which can thermalize ions that can become trapped by a negatively charged dust grain and reduce the dust grain charge. Since the collision enhanced current depends on the ion motion, there are two models that incorporate it, one for the bulk~\cite{Khrapak2005} where the ion velocity is low and one for the sheath~\cite{Piel2015} where the ion velocity is significant. The work by \citet{Piel2015} showed the effects of the collision enhanced current on the surface potential of a single dust grain, the current work examines the effects on a dust crystal.

Modeling efforts through a particle description of dust have been more interested in waves or voids \cite{Schwabe2013,Schwabe2014} than in the structure of the plasma crystal. These studies either significantly simplified the problem to allow for tracking a large number of dust grains \cite{Schwabe2013} or examined smaller crystals with fewer simplifications \cite{Davoudabadi2007}.
An example of a simplified model of a large crystal is the work by \citet{Klumov2007} which examined a crystallization wave. In this work, the plasma profile was prescribed, not computed, the ion wake was neglected and the dust grain charge was specified which allowed for examining a crystal containing 900 dust grains.
The other approach is demonstrated by the work of \citet{Davoudabadi2007} which examined small dust crystals, composed of 4-19 grains, in a cylindrical plasma reactor. The plasma was modeled using a continuum approach via a local field approximation for the plasma, the ion wake was modeled as a dipole, and the dust grain charge was computed as a function of time for each individual grain. The study performed a qualitative comparison of the crystal structure against the experimental data, however due to differences in the reactor conditions, quantitative comparisons were not be performed.

The present work is a computational study examining dust crystals in a GEC cell, a setup previously investigated experimentally \cite{Thomas2005,Hartmann2014}. 
The GEC cell was first proposed to facilitate comparisons of experimental plasma studies done by different groups \cite{Boeuf1995} and its popularity with the plasma community~\cite{Kong2016,Siari2020} led to it being a common choice for complex plasma research.
The reactor cross-section is shaped like an `H' and the electrodes do not extend radially to the outer edge of the reactor. The shape of the GEC cell reactor results in a strong radial dependence with a maximum plasma density that is not on the reactor centerline \cite{Land2009}. 
While GEC cells are commonly investigated in experiments, they are less common in computational studies. Models often treat the reactor as either a one-dimensional parallel plate setup \cite{Davoudabadi2006} or a cylinder\cite{Davoudabadi2007} which complicates comparing experimental and computational results. 

The present study models a GEC cell dust experiment performed by \citet{Douglass2012} to gain further understanding of the dust crystal in this experiment. Further,  the dust charging was separately represented by the OML approach and the collision enhanced charging (CEC) approach \cite{Piel2015} to quantify the effect of the collision enhanced charge on the crystal. % and determine whether the CEC model improves the agreement with the experimental results.}
The plasma is modeled using a continuum approach illustrated in \S\ref{sec:AllEqs}. 
To model dust dynamics, dust grains are individually tracked using equations described in \S\ref{sec:AllEqs}. 
The numerical methods  are illustrated in \S \ref{sec:NumMethod}.
The results are discussed in \S\ref{sec:Results} and conclusions are made in \S \ref{sec:Conclusions}.

\section{\label{sec:AllEqs}Governing Equations}  

\subsection{\label{sec:PlasmaEqs}Plasma Equations}%%% Plasma model
The plasma model, which was presented in detail and validated previously \cite{Wilson2018}, is based on a continuum description using a second-order local mean energy approximation\cite{Gogolides1992}.
In this model, the equations governing the electron number density $n_e$, ion number density $n_i$, and electron energy density $\omega _e$ are:
\begin{equation} \label{eq:ne,i}
\frac{\partial n_{e(i)}}{\partial t} +\bm{\nabla}\cdot\bm{\Gamma}_{e(i)}=S_{e(i)} ,
\end{equation}
\begin{equation} \label{eq:EnergyTransport}
\frac{\partial\omega _e}{\partial t}+\bm{\nabla}\cdot\bm{\Gamma}_\omega =-e\bm{\Gamma}_e\cdot\bm{E}+S_\omega ,
\end{equation}
where
\begin{equation} \label{eq:neFlux}
\bm{\Gamma}_{e(i)}=\text{sgn}(q)n_{e(i)}\mu_{e(i)}\bm{E}-D_{e(i)}\bm{\nabla}n_{e(i)}  ,
\end{equation}
\begin{equation} \label{eq:EnergyFlux}
\bm{\Gamma}_\omega =\frac{5}{3}\left(-\omega_e\mu_e\bm{E}-D_e\bm{\nabla}\omega_e\right) ,
\end{equation}
and $\text{sgn}(q)$ is $1$ for ions and $-1$ for electrons and electron energy.
Here,  $\bm{E}$ is the electric field calculated by:
\begin{equation} \label{eq:E_inst}
\bm{E}=-\bm{\nabla}\phi ,
\end{equation}
where $\phi$ is the electric potential which satisfies Poisson's equation:
\begin{equation} \label{eq:Poissons}
\nabla^2\phi=\frac{e}{\epsilon_0}\left( n_e-n_i\right),
\end{equation}

\noindent
where $e$ is the electron charge. 
In the equations above, $\mu_{e(i)}$ is the electron (ion) mobility, $D_{e(i)}$ is the electron (ion) diffusion coefficient and $\bm{E}$ is the electric field. 
Eqs. (\ref{eq:neFlux}-\ref{eq:EnergyFlux}) define the fluxes of  electrons (ions) and energy, respectively. 
The source term $S_{e(i)}$ in Eq. (\ref{eq:ne,i}) accounts for the electrons and ions created by ionization. 
The gas is assumed to be singly ionized, therefore $S_i=S_e=k_in_en_\text{gas}$ where $k_i$ is the ionization rate coefficient. 
The ionization rate was determined by BOLSIG+ \cite{Hagelaar2005} which solves the electron Boltzmann equation and tabulates the ionization rate and excitation rates as a function of the mean electron energy. 
The mean electron energy $\varepsilon$, the electron temperature $T_e$ and  $\omega _e$ are  correlated  through
\begin{equation}
\omega_e =n_e\varepsilon=\frac{3}{2}k_B n_e T_e, 
\end{equation}
where $k_B$ is the Boltzmann constant. 
In the energy equation, Eq. (\ref{eq:EnergyTransport}), the term $-e\bm{\Gamma}_e\cdot\bm{E}$  accounts for the ohmic or joule heating of the electrons in the electric field and the term $S_\omega=S_e H_i$ accounts for the energy loss due to ionization and excitation, where $H_i$ is the ionization energy.
Since this study examines small mono-layer dust crystals containing 40 - 100 dust grains, the dust loading is low enough that the dust does not significantly affect the plasma \cite{Havnes1990} and the plasma can be modeled without accounting for the modification of plasma by dust. 

The boundary conditions for the plasma model are an applied voltage on the lower electrode with the remaining surfaces grounded. For the powered electrode, the electric potential is given by:
\begin{equation} 
\phi=V_\text{DC}+V_\text{RF}\sin\left(2\pi f t\right)  ,
\end{equation}
where $V_\text{DC}$ is the direct current voltage, $V_\text{RF}$ is the radio-frequency voltage and $f$ is the RF frequency. 
For the reactor geometry considered here, there is a difference in the area of the powered electrode and the total grounded area, which causes a natural DC bias \cite{Passchier1993,Land2009}. The bias in  the model was set to match the bias measured in the experiments~\cite{Douglass2012}.
The number and energy density boundary conditions assume that all surfaces are perfectly absorbing with no reflection \cite{Passchier1993,Davoudabadi2009}, therefore, the ion and electron number density at the surface are zero, i.e., $n_i=n_e=0$, and correspondingly, $\omega_e=0$. 
Modeling the electron drift, thermal, and diffusion fluxes to the surface produces a slightly higher plasma density \cite{Wilson2018} however the simpler boundary condition is used here to enable comparisons with the work of \citet{Douglass2012}.

\subsection{\label{sec:DustEqs}Dust Equations}%%% Dust model
Once the plasma reaches the quasi-steady state, the plasma values are averaged over one RF cycle and the dust grains are added. Because the dust loading is very low here, the plasma is assumed not affected by the dust; hence, it is not updated during the dust simulation.

%DUST LOCATION
Dust grain motion is fully three-dimensional with the position, velocity, and charge evolving over time. The equations for the position and velocity vectors of each grain are:
\begin{equation} \label{eq:drdt}
\frac{d\bm{r}_p}{dt}=\bm{v}_{p}
\end{equation}
and
\begin{equation} \label{eq:dvdt}
\frac{d\bm{v}_{p}}{dt}=\frac{\bm{F}_{t}}{m_p},
\end{equation}
where $\bm{F}_t$ is the total force on the grain and $m_p$ is the mass of the dust grain. The specific forces experienced by the dust particle is illustrated in sec.~\ref{subsub:Forces}.

%Dust charge

The charge of the dust grain $Q_p$ varies with time and is determined by the electron and ion currents to the grain, denoted by $I_e$ and $I_i$, respectively:
\begin{equation} \label{eq:dQ}
\frac{d Q_{p}}{dt}=I_e+I_i,
\end{equation}
The dust charge and the floating potential of the dust grain, $(\phi_p-\phi)$, are related by:
\begin{equation}
Q_p = 4\pi\epsilon_0 a_p (\phi_p-\phi).
\end{equation}
where $a_p$ is the grain radius. Two different charging approaches are used in the present study, which are discussed in the following two subsections. For the plasma regions of interest here, the modified OML model using a shifted Maxwellian for the ions produces the same results as the standard OML model using a monoenergetic beam, so the shifted Maxwellian variation is not considered separately. 

\subsubsection{Orbital-motion limited  (OML) charging}
The classic approach to determine the ion and electron currents is the OML theory \cite{Allen1992} which assumes a Maxwellian distribution for the electrons and a monoenergetic beam for the ions \cite{Barnes1992}. The electron current is written in terms of the floating potential as:
\begin{equation} \label{eq:electronCurrent}
I_e = -e n_e\pi a_p^2 v_{the}\exp\left[\frac{e(\phi_p-\phi)}{k_B T_e}\right], 
\end{equation}
where $a_p$ is the radius of the dust grain and $v_{the}=\sqrt{8k_BT_e/\pi m_e}$ is the electron thermal velocity. 
The ion current is defined as: 
\begin{equation}
I_i = e n_i\pi a_p^2 v_s\left[1-\frac{2e(\phi_p-\phi)}{m_i v_s^2}\right].
\label{eq:ionCurrent}
\end{equation}
The velocity used to obtain the ion current is the ion mean speed toward the dust grain, $v_s=\sqrt{v_{thi}^2+v_{ip}^2}$, where $v_{thi}=\sqrt{8k_BT_i/\pi m_i}$ is the ion thermal velocity and the  $\bm{v}_{ip}=\bm{v}_i-\bm{v}_p$ is the drift motion velocity relative to the dust grain.

\subsubsection{Collision Enhanced Charging (CEC)}
The second charging method examined here is the collision enhanced ion current model proposed by \citet{Piel2015}.
In this model, the ion current includes the effect of collisions which can cause a flowing ion to become a thermal ion. If the resulting potential energy is below a certain threshold, the ion will become trapped and collected by the negatively charged dust grain which can substantially reduce the dust charge compared to the collisionless OML model. 
In CEC, the electron current to the dust grain is the same as in the OML model, viz. Eq.~(\ref{eq:electronCurrent}).

The ion current is the sum of the collisionless current $I_i^\mathrm{(s)}$ and the collision enhanced current $I_i^\mathrm{(cs)}$:
\begin{equation} \label{eq:CEC_Istandard}
    I_i^\mathrm{(s)}=e n_i \pi a_p^2 v_s \left(1-\frac{2 e (\phi_p-\phi)}{m_i v_s^2}\right)\left(1-\frac{Z^*}{l_i}\right)
\end{equation}
\begin{equation} \label{eq:CEC_I}
    I_i^\mathrm{(cs)}=e n_i \pi v_s \frac{{R^*}^2 Z^*}{l_i}
\end{equation} 
where $v_s$ is the ion mean speed toward the dust grain, and $l_i=7.0\mathrm{mm}/P_{gas}\mathrm{(Pa)}$ is the ion mean free path, which is based on the charge exchange cross-section of argon ions~\cite{Phelps1991}. 
The terms $Z^*$ and $R^*$ and the major semi-axis and the minor semi-axis, respectively, of the ellipse that defines the trapping region, the calculation of these variables is illustrated in Appendix \ref{sec:R*Z*Derive}.  It is noted that the total ion current, the summation of Eq.~(\ref{eq:CEC_Istandard}) and Eq.~(\ref{eq:CEC_I}), reduces to the OML model if $Z^*$ is zero.

%DUST Forces
\subsubsection{Dust Forces}
\label{subsub:Forces}
The forces experienced by the dust grain are: the gravitational force, the electric force, the ion drag force, the neutral drag force, and the force due to interactions with other dust grains.

%Gravity
%\subsubsection{Gravitational force}
The force on the dust grain due to gravity is given by
$ \bm{F}_g = -m_p g \bm{i}_z $
where $\bm{i}_z$ is the unit vector in the vertical direction.

%Electric force
%\subsubsection{Electric force}
The electric force is defined as:
\begin{equation}
\bm{F}_e = Q_p\bm{E}\left[1+\frac{\left(\frac{a_p}{\lambda_D}\right)^2}{3\left(1+\frac{a_p}{\lambda_D}\right)}\right],
\label{eq:Fe}
\end{equation}
where  
\begin{equation} \label{eq:lin_debye}
\lambda_D = \left(\frac{e^2 n_e}{\epsilon_0 k_B T_e}+\frac{e^2 n_i}{\epsilon_0 k_B T_e+m_i\left(v_{ip}\right)^2}\right)^{-1/2},
\end{equation}
is the effective (linearized) Debye length at the grain location.
The second term in the brackets in Eq.~(\ref{eq:Fe}) accounts for the fact that the grain is not a geometric point charge. 

%Ion drag force
%\subsubsection{Ion drag force} 
The ion drag force accounts for the effect of ions that are collected by the grain as well as ions that are scattered in the field of the grain. 
The ion drag force is calculated by \cite{Barnes1992}:
\begin{equation} \label{eq:Fid}
\bm{F}_{id} = \pi n_i v_s m_i \bm{v}_{ip}\left(b_c\right)^2 +4\pi n_i v_s m_i \bm{v}_{ip}\left(b_{\pi/2}\right)^2 \ln\Gamma ,
\end{equation}
where the first term accounts for the ions collected by the grain with the collection impact parameter defined as:
\begin{equation}
b_c=a_p\left[1-\frac{2e(\phi_p-\phi)}{m_i(v_s)^2}\right]^{1/2} ,
\end{equation}
and the second, orbital, term accounts for the ions that are scattered in the field of the grain due to Coulomb collisions. The orbital impact parameter corresponding to a $90^\circ$ deflection is:
\begin{equation}
b_{\pi/2}=a_p\frac{e(\phi_p-\phi)}{m_i v_s^2} .
\end{equation}
The Coulomb logarithm is given by:
\begin{equation}
\ln\Gamma = \frac{1}{2}\ln\left[\frac{(\lambda_{D_e})^2+(b_{\pi/2})^2}{(b_c)^2+(b_{\pi/2})^2}\right] ,
\end{equation}
where $\lambda_{D_e}=\sqrt{\epsilon_0 k_B T_e/e^2 n_e}$ is the electron Debye length. It is noted that \citet{Hutchinson2006} showed that  the accuracy of the ion drag force model of \citet{Barnes1992} may not be sufficient for an ion Mach number ranging between 1 and 2. As will be seen, the dust crystals simulated here levitate near the edge of the sheath where the ion Mach number is significantly larger; and hence, \citet{Barnes1992}'s model is adequate. 

%neutral drag force
%\subsubsection{Neutral drag force}
While there is no neutral gas flow, the neutral drag force has a damping effect on dust grain motion \cite{Nitter1996} so it is included here. 
The model for the neutral drag force assumes specular reflection of the neutral particles after collision with the dust grain. 
The force is defined by the Epstein relation:
\begin{equation}
\bm{F}_{nd}^m=\frac{8}{3}\sqrt{2\pi}(a_p^m)^2 m_n n_n v_{th_n} (\bm{v}_n^m-\bm{v}_p^m) .
\end{equation}
The neutral temperature is constant and equal to the ion temperature, $T_n=T_i$, and the neutral density is $n_n=P/k_B T_n$. 

Since the model assumes that the dust does not affect the plasma, the ion wake that occurs around the negatively charged dust grains is modeled in the interaction force instead of in the plasma profile. 
There are a variety of models proposed for the ion wake ranging from detailed models that include collisions and non-Maxwellian ions~\cite{Kompaneets2016} to simple models which treat the particle and wake as a dipole~\cite{Davoudabadi2007} or a pair of particles~\cite{Qiao2010}. Since the most significant effect of the ion wake is the vertical alignment \cite{Samarian2005} seen between layers in a dust crystal and this study only examines mono-layer crystals, the simple image particle model is used for the ion wake.
Using this model means effects of the ion wake on crystal parameters, such as the crystal stability, can not be examined in detail, however since the ion wake is not the focus of this work, the image particle model is adequate.
In the image particle model, the dust grain and ion wake are modeled by including a positively charged image particle downstream of the dust grain \cite{Qiao2010}. 
The image particle has a charge $q$ and is located at a distance $l$ downstream of the dust grain, where `downstream' is defined based on the direction of the ion flow at the location of the dust grain. Here, the values of these two parameters are set to $q=2 \lvert Q\rvert /3$ and $l=2 \lambda_D/3$, as suggested by \citet{Qiao2010}.

The force on grain $m$ due to grain $n$ is a function of the potential around particle $n$, which is indicated by $\phi_n$, and the grain separation distance $\bm{r}_{mn}=\bm{r}_m-\bm{r}_n$. 
The potential around the dust grain is modeled using the screened Coulomb potential since the particle separation distance is comparable to the plasma screening length.
Since the plasma screening length has been experimentally measured to be on the same order as the bulk electron Debye length \cite{Konopka2000,Fortov2005} but is sometimes attributed to the ions \cite{Carstensen2010,Kompaneets2007}, this study uses the linearized Debye length at the grain location, given by Eq. (\ref{eq:lin_debye}), as the screening length. 

%\paragraph{Screened Coulomb potential} 
The negatively charged dust grain will attract the positive ions which results in a reduction of the apparent charge viewed from a long distance \cite{Laframboise1966}.
This effect is referred to as plasma screening and is modeled using the screened Coulomb potential where
the potential around grain $n$ is given by:
\begin{equation}
\phi_{Qn}=\frac{1}{4\pi\epsilon_0}\frac{Q_n}{r}\exp\left(\frac{-r}{\lambda_D^{(n)}}\right) ,
\label{eq:screenedCoulomb}
\end{equation}
and the potential around the image particle associated with grain $n$ is given by:
\begin{equation}
\phi_{qn}=\frac{1}{4\pi\epsilon_0}\frac{q_n}{r}\exp\left(\frac{-r}{\lambda_D^{(n)}}\right) .
\end{equation}
The force on grain $m$ due to grain $n$ is the summation of the force due to the grain itself and the force due to the image particle:
\begin{equation} \begin{gathered}
\bm{F}_{mn}=\frac{1}{4\pi\epsilon_0}\left(1+\frac{r_{mn}}{\lambda_D^{(n)}}\right)\exp\left(\frac{-r_{mn}}{\lambda_D^{(n)}}\right)\left[\frac{Q_m Q_n}{r_{mn}^2}\right]\bm{i}_{mn} \\
+\frac{1}{4\pi\epsilon_0}\left(1+\frac{r_{mq}}{\lambda_D^{(n)}}\right)\exp\left(\frac{-r_{mq}}{\lambda_D^{(n)}}\right)\left[\frac{Q_m q_n}{r_{mq}^2}\right]\bm{i}_{mq} ,
\end{gathered} \end{equation}
where $r_{mn}$ is the distance between grain $m$ and grain $n$, $\bm{i}_{mn}$ is the unit vector from grain $m$ to grain $n$, and $r_{mq}$ and $\bm{i}_{mq}$ are the distance and unit vector, respectively, between grain $m$ and the image particle associated with grain $n$. The force on grain $m$ due to the ion wake that forms around it is included in the ion drag force and not the interaction force.

\section{\label{sec:NumMethod}Numerical Method}  %%% Numerical method

Details of the numerical methods used for the plasma simulation were provided in a previous work\cite{Wilson2018} with a summary  here.
The time derivatives in the plasma equations use a second order scheme and the equations are iterated until the system is solved simultaneously.
The spatial discretization of the fluxes uses the Scharfetter-Gummel scheme~\cite{Scharfetter1969,Boeuf1987,Hagelaar2000}.
The remaining spatial discretizations use a second-order finite difference method applied in an axisymmetric cylindrical coordinate consistent with the axisymmetric geometry of the plasma reactor.
A uniform, staggered grid is used with the primary variables, $\phi$, $n_{e,i}$, and $\omega_e$, evaluated on the nodes, and the electric field $\bm{E}$ and the fluxes  $\bm{\Gamma}_{e,i,\omega}$ evaluated halfway between the nodes. The boundary passes through the nodes.

Once the plasma has reached a quasi-steady state, i.e., the time-averaged plasma properties  do not change with time, the plasma properties are averaged over one RF period and used as an input to the dust model.
The individual dust grains are tracked in a three dimensional Lagrangian framework with no axisymmetric assumption.
A linear interpolation in both the $r$ and $z$ directions is used to obtain the plasma values at the location of each grain and the forces on the grains are determined from the interpolated plasma values. 
The interaction between dust grains is computed for pairs of grains based on the current location of each dust grain and the local plasma Debye length.
Then, the dust  Eqs. (\ref{eq:drdt}-\ref{eq:dQ}) are solved using the second-order Adams-Bashforth scheme to advance to the next time step.

\section{\label{sec:Results}Results and Discussion}  %%% Results

The configuration of the reactor studied here resembles a GEC cell experimentally investigated in a previous work \cite{Douglass2012}. 
The reactor was 22.81 cm high with a 2.54 cm gap between the electrodes. 
The reactor and electrode radii were 12.7 cm and  5.4 cm, respectively. 
The lower electrode had a 1.27 cm radius circular cutout with a 1mm depth, needed to create a parabolic profile~\cite{Klumov2019} in the radial electric field to confine the dust to the center of the reactor. 
The operating parameters used in the simulation are given in Table \ref{tab:parameters}.
Cases were simulated with 40, 60, 80, and 100 dust grains. 
Three sizes of dust grains were examined, with diameters of 6.37, 8.89  and 11.93 $\mu \mathrm{m}$.

\begin{figure}[h]   
  \centering
  \begin{subfigure}[ht]{0.45\textwidth}
    \centering
    \includegraphics[scale=0.3]{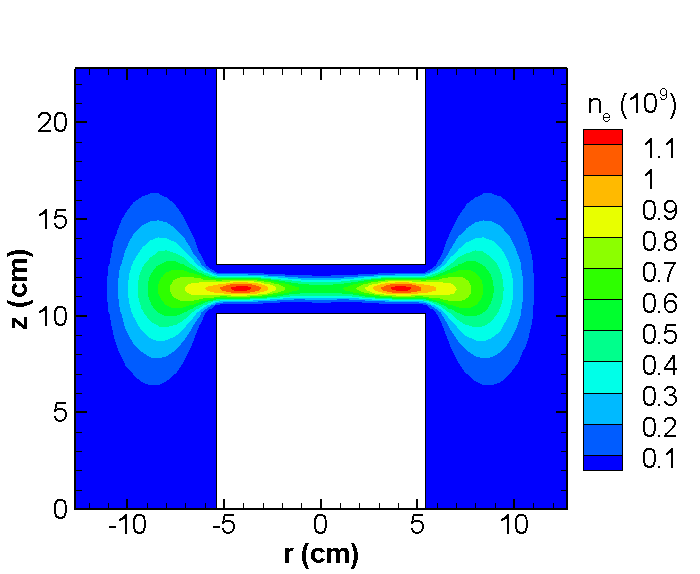}
    \caption{}
    \label{fig:ne_T}
  \end{subfigure}
  ~
  \begin{subfigure}[ht]{0.45\textwidth}
    \centering
    \includegraphics[scale=0.3]{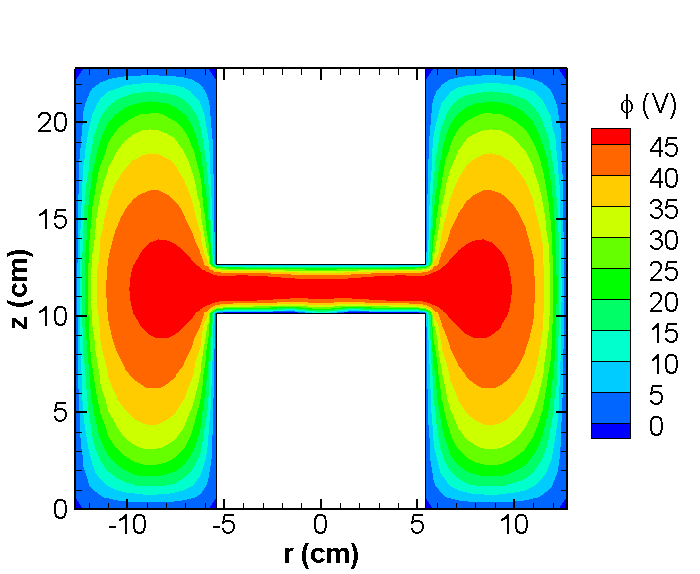}
    \caption{}
    \label{fig:potential_T}
  \end{subfigure}
  \caption{Contour plots of RF-averaged plasma variables for a 150 mTorr argon plasma with an RF voltage of 60V and a DC bias of -5V, cutout radius is 1.27cm: (a) electron number density (in cm$^{-3}$); and (b) electric potential (in V).} 
  \label{fig:2D_full}
\end{figure}

Figure \ref{fig:2D_full} shows the contours of electron number density and electric potential for the case with an RF voltage of 60 V.
The geometric form  of the reactor including side regions past the electrodes causes the plasma to expand radially and as a result, the maximum plasma density to occur away from the reactor centerline, viz. $r\sim 5\mathrm{cm}$, as seen in Fig. \ref{fig:2D_full}(a). 
The specific role of the electrode cutout is to confine the dust crystal radially by slightly altering the electric field in the radial direction. Hence, the region of interest for the dust crystal, as will be seen, is the section of the reactor between the electrodes with a radius less than the radius of the cutout ($r=1.27$cm). 

\begin{figure}%[h]   
  \centering
  \begin{subfigure}[t]{0.315\textwidth}
    \centering
    \includegraphics[scale=0.3]{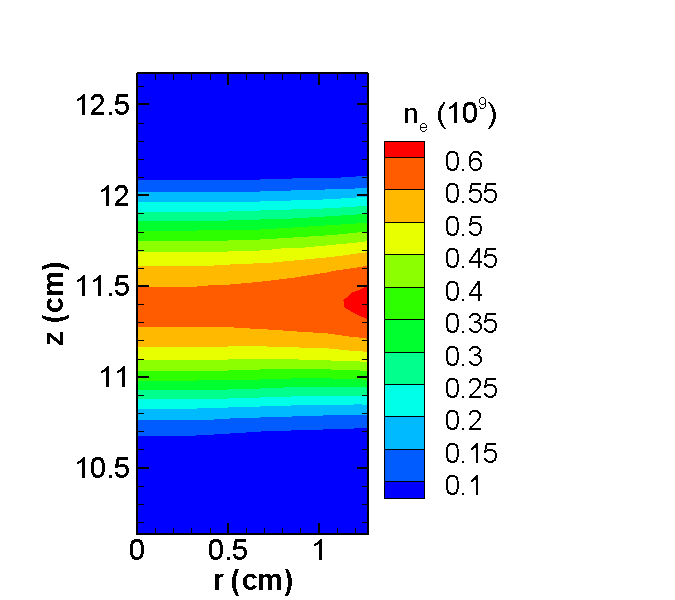}
    \caption{}
    \label{fig:ne2}
  \end{subfigure}
  ~
  \begin{subfigure}[t]{0.315\textwidth}
    \centering
    \includegraphics[scale=0.3]{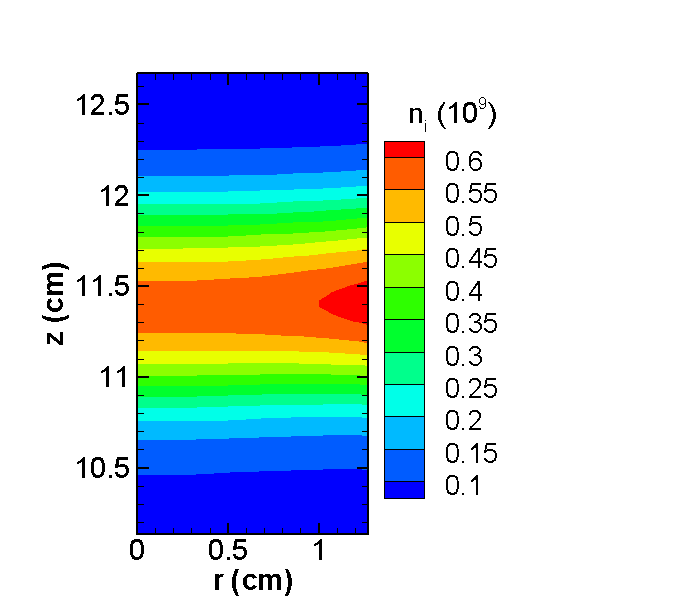}
    \caption{}
    \label{fig:ni2}
  \end{subfigure}
  ~
  \begin{subfigure}[t]{0.315\textwidth}
    \centering
    \includegraphics[scale=0.3]{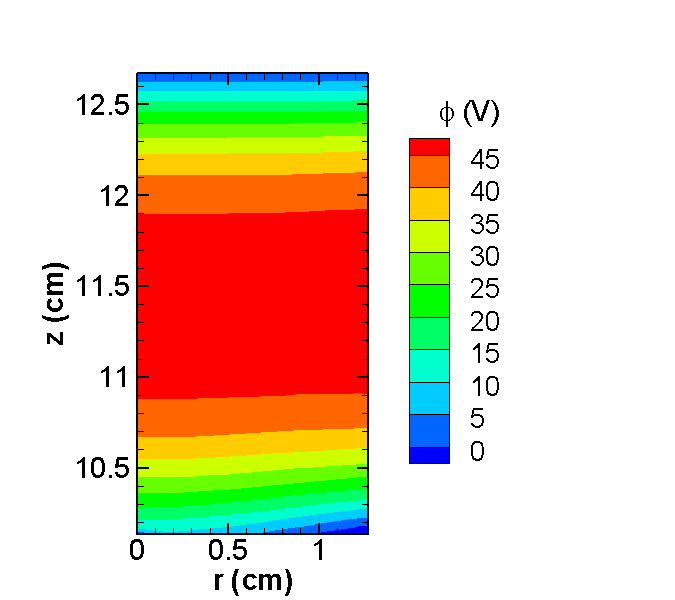}
    \caption{}
    \label{fig:Potential2}
  \end{subfigure}
  
  \begin{subfigure}[t]{0.315\textwidth} 
    \centering
    \includegraphics[scale=0.3]{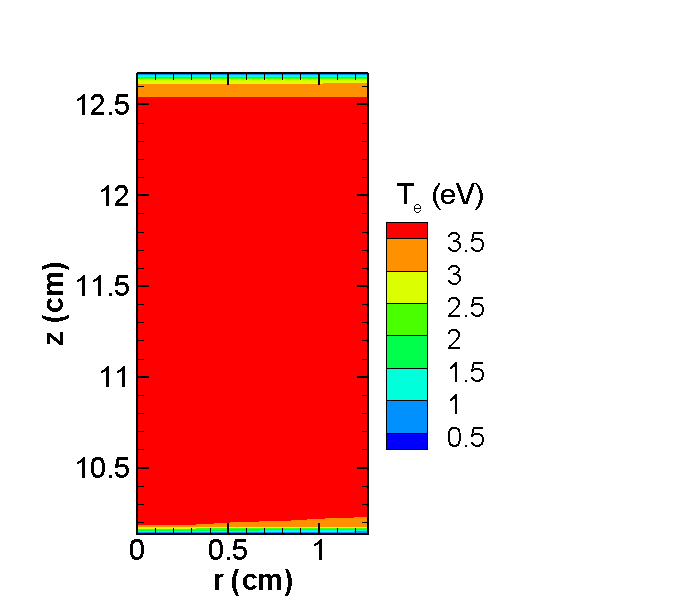}
    \caption{}
    \label{fig:Te2}
  \end{subfigure}
  ~
  \begin{subfigure}[t]{0.315\textwidth}  
    \centering
    \includegraphics[scale=0.3]{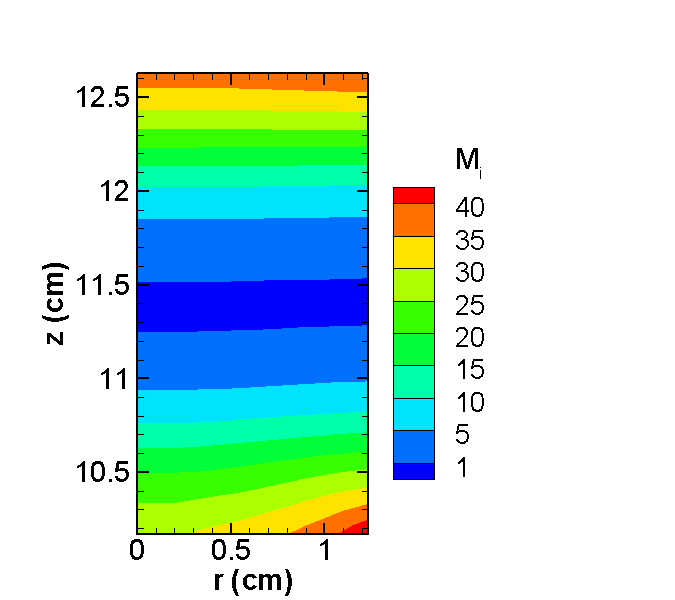}
    \caption{}
    \label{fig:Mion}
  \end{subfigure}
  ~
  \begin{subfigure}[t]{0.315\textwidth}  
    \centering
    \includegraphics[scale=0.3]{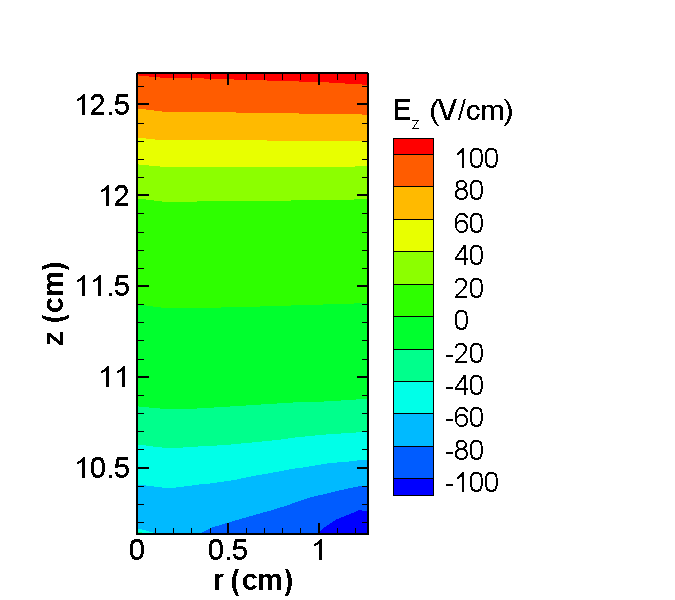}
    \caption{}
    \label{fig:Ez2}
  \end{subfigure}
  \caption{Spatial variation of RF-averaged plasma variables in the region between the electrodes, the right edge is the location of the electrode cutout, the top and bottom edges are the electrodes located at $z=12.675$ and $10.135$cm, respectively; (a) electron number density (in cm$^{-3}$); (b) ion number density (in cm$^{-3}$); (c) electric potential (in V); (d) electron temperature (in eV); (e) ion Mach number; and (f) vertical electric field in (V/cm).}%; and (f) radial electric field (in V/cm).} 
  \label{fig:2D_zoom}
\end{figure}

Figure \ref{fig:2D_zoom} shows the contour plots of plasma variables in the region of interest. This is the region where the dust crystal is confined.  
The maximum plasma density in the reactor occurs outside this region, as seen in Fig.~\ref{fig:2D_full}(a), and the maximum plasma density within the region of interest is about half of that in the entire reactor.
Both electric potential and electric field, shown in Fig.~\ref{fig:2D_zoom}(c) and (f), respectively,  have a radial dependence along the lower electrode because of the electrode cutout. 
Based on correspondence with the authors of \citet{Douglass2012}, a parabolic voltage distribution $V_\mathrm{cut}=15-15({r}/{r_c})^2$ is added to the applied RF voltage to model this cutout on the electrode in the radial direction.
The ion Mach number is shown in Fig.~\ref{fig:2D_zoom}(e). 
Due to the assumption that the ion temperature $T_i$ is constant, the ion thermal velocity is constant. The ion drift velocity is obtained from the ion flux by $v_i=\Gamma_i/n_i$. Since the ion flux is a function of the electric field, the ion drift velocity is also dependent on the electric field. 
The ion Mach number is supersonic over most of the region with the exception of a narrow band in the bulk of the plasma where the electric field is close to zero. The subsonic band is 2.5 mm in height and is located 11 mm above the lower electrode. 
Since the OML model assumes a monoenergetic beam for the ions and the CEC model assumes a significant ion velocity, neither model is valid for the region in the bulk where the ion velocity is low. Therefore, the dust results are only displayed for the region in the lower sheath below 11.2cm in the vertical direction, which is within 10mm above the lower electrode.

\begin{figure}%[h]   
  \centering
  \begin{subfigure}[t]{0.315\textwidth}
    \centering
    \includegraphics[scale=0.25]{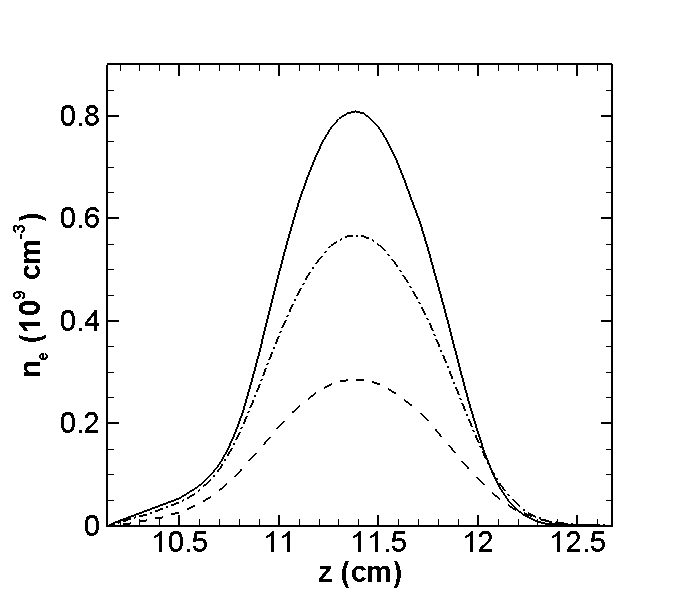}
    \caption{}
    \label{fig:ne_axis}
  \end{subfigure}
  ~
  \begin{subfigure}[t]{0.315\textwidth}
    \centering
    \includegraphics[scale=0.25]{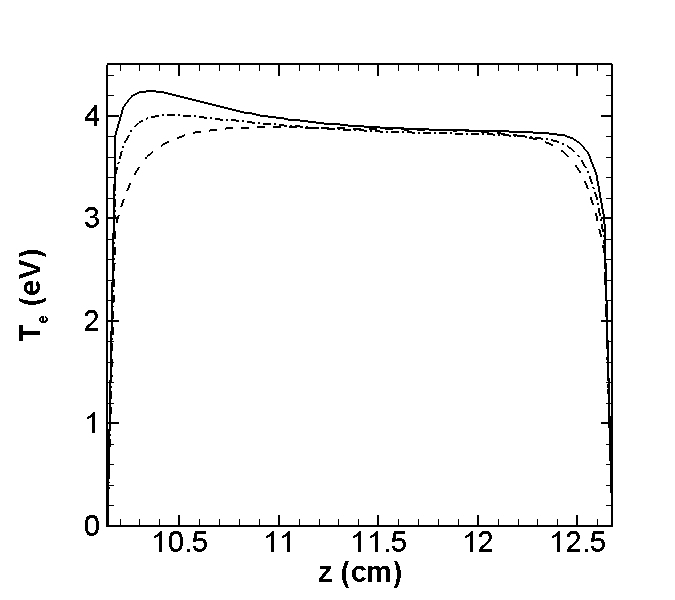}
    \caption{}
    \label{fig:Te_axis}
  \end{subfigure}
  ~
  \begin{subfigure}[t]{0.315\textwidth}
    \centering
    \includegraphics[scale=0.25]{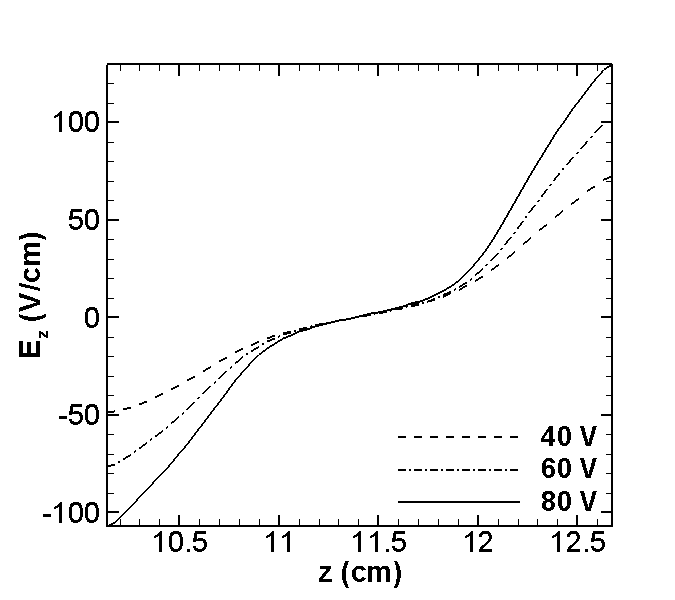}
    \caption{}
    \label{fig:Ez_axis}
  \end{subfigure}
  \caption{Plasma profiles along $r=0$: (a) electron number density (in cm$^{-3}$); (b) electron temperature (in eV); and (c) vertical electric field in (V/cm).} 
  \label{fig:axis_profiles}
\end{figure}

Figure \ref{fig:axis_profiles} shows the effect of changing the RF voltage on the plasma.
The number density in the bulk is a function of the RF voltage. For this setup, the maximum number density increased by a factor of four as the RF voltage was doubled from 40 to 80 V. 
The electron temperature was less sensitive to the RF voltage; the only effect was a small increase in the electron temperature in the sheath as the RF voltage increased. 
The electric field did not show an effect in the bulk however the electric field in the sheath doubled when the RF voltage doubled.
There is an asymmetry in the plasma due to the constant DC voltage of -5 V, this is most obvious in the temperature profile and can also be seen in the larger sheath near the upper electrode and the higher electric field there.

The experimental data of \citet{Douglass2012} was used to validate the dust model. In the experiments, the trajectory of a single grain was recorded and the electric force was determined from the acceleration using an Epstein drag coefficient to calculate the drag force on the dust grain. 
In the current modeling study, a single 8.89 $\mu \mathrm{m}$ diameter dust grain was released along the reactor centerline, $r=0$; the grain was tracked and the computed electric force was compared against the experimental data \cite{Douglass2012}. This validation exercise was conduced for both the OML and CEC charging models.

\begin{figure}[h]   
  \centering
  \begin{subfigure}[t]{0.315\textwidth}
    \centering
    \includegraphics[scale=0.19]{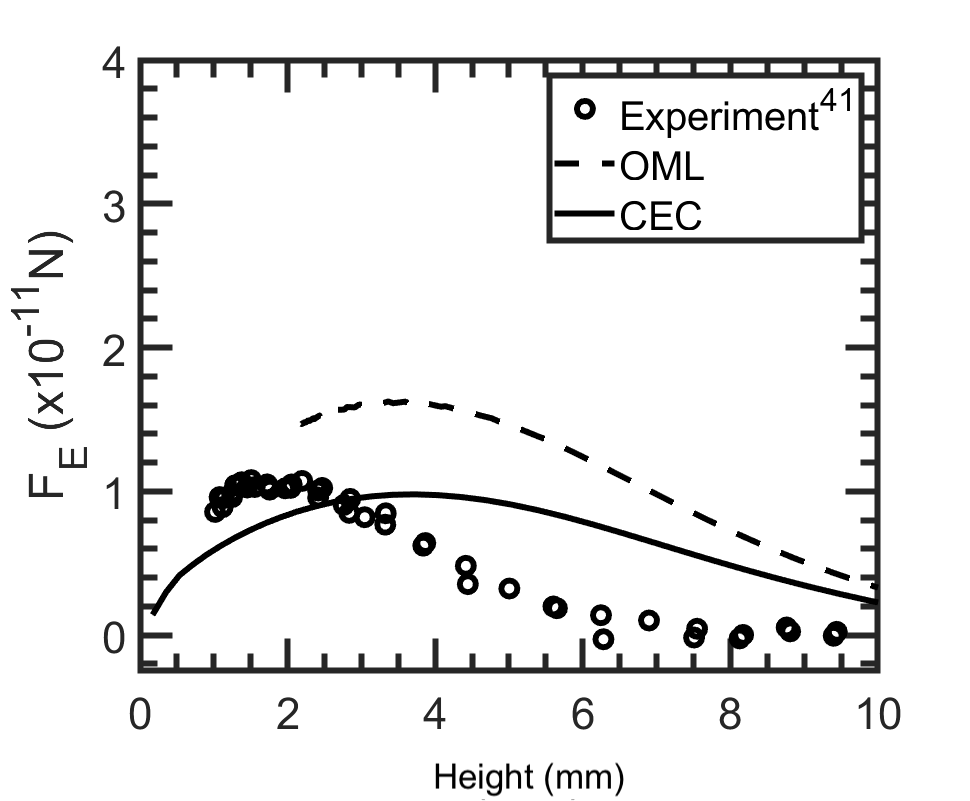}
    \caption{}
    \label{fig:30V_Fe}
  \end{subfigure}
  ~
  \begin{subfigure}[t]{0.315\textwidth}
    \centering
    \includegraphics[scale=0.19]{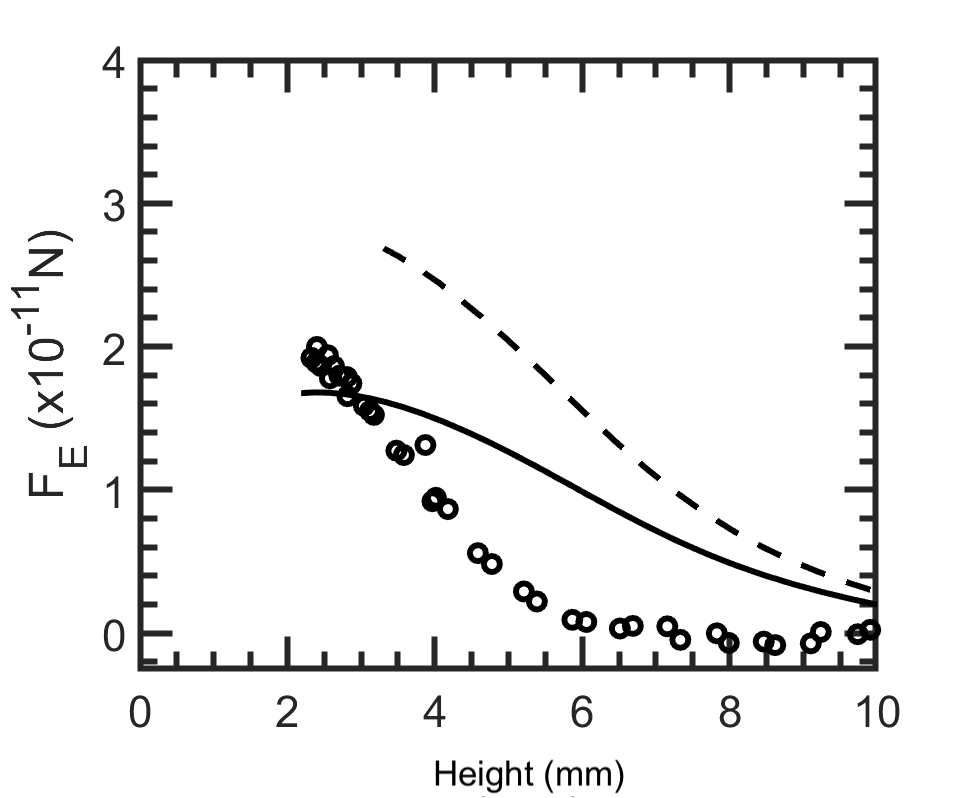}
    \caption{}
    \label{fig:47V_Fe}
  \end{subfigure}
  ~
  \begin{subfigure}[t]{0.315\textwidth}
    \centering
    \includegraphics[scale=0.19]{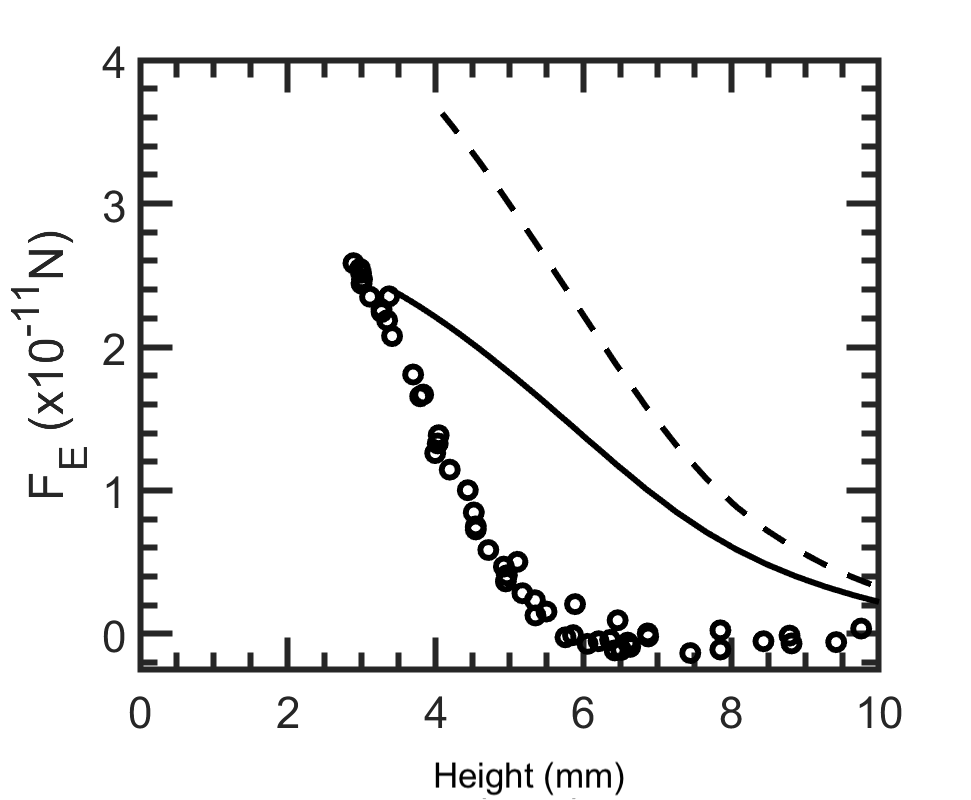}
    \caption{}
    \label{fig:64V_Fe}
  \end{subfigure}
  \caption{Electric force on a 8.89 $\mu$m dust grain vs height above the electrode for experiments\cite{Douglass2012} (symbols), OML model (dashed lines), and CEC model (solid line); (a) 30 V plasma; (b) 47 V;  (c) 64 V.} 
  \label{fig:Dust_Fe}
\end{figure}

Figure \ref{fig:Dust_Fe} shows the electric force profiles from the current simulations based on OML and CEC charging models as well as the previous experimental data from \citet{Douglass2012}.
The electric force in the experiment is almost constant at heights above 6mm and increases rapidly as the height decreases suggesting a transition between the bulk and sheath around 6mm.
In contrast, neither the OML nor CEC models show an obvious transition point between bulk and sheath. 
It is noted that  \citet{Douglass2012} did not directly measure the electric force  but used the measured data in combination with a theoretical equation for the grain motion where the ion drag force was completely neglected. In contrast, the ion drag force is included in our model. Nonetheless,  the trends of all three curves are similar in this figure but it is evident that between two models, CEC does a better job in terms of the error, defined as the difference between the modeling and experimental electric forces, for all range of $z$ and all considered grain sizes.

\begin{figure}%[h]   
  \centering
  \begin{subfigure}[t]{0.315\textwidth}
    \centering
    \includegraphics[scale=0.19]{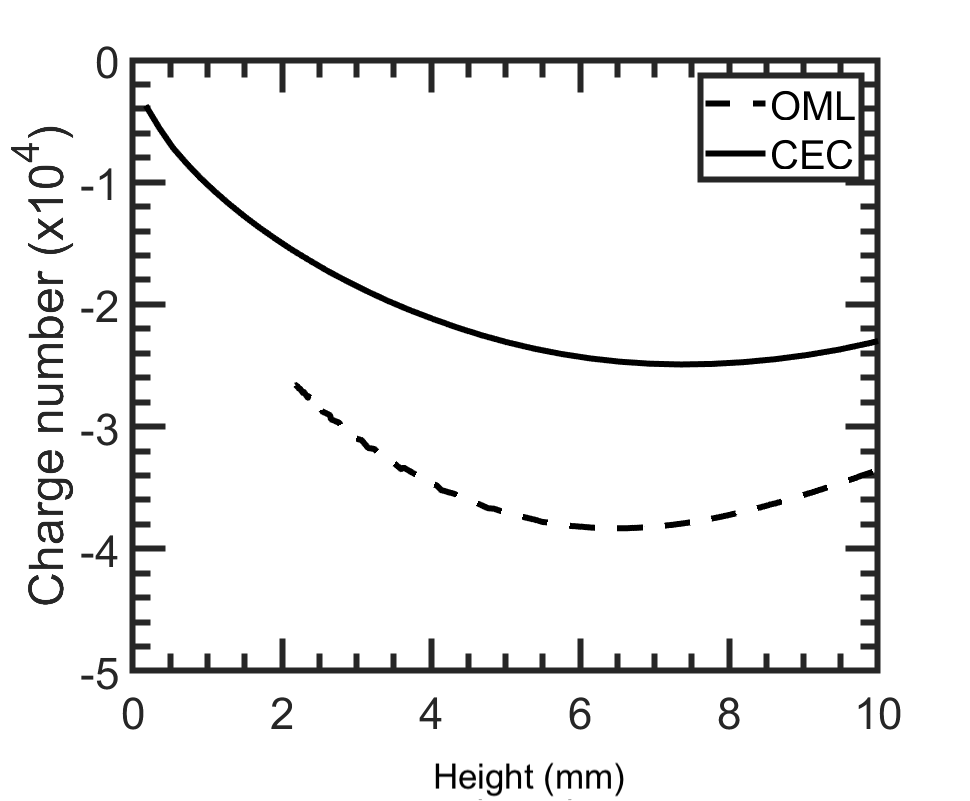}
    \caption{}
    \label{fig:30V_Q}
  \end{subfigure}
  ~
  \begin{subfigure}[t]{0.315\textwidth}
    \centering
    \includegraphics[scale=0.19]{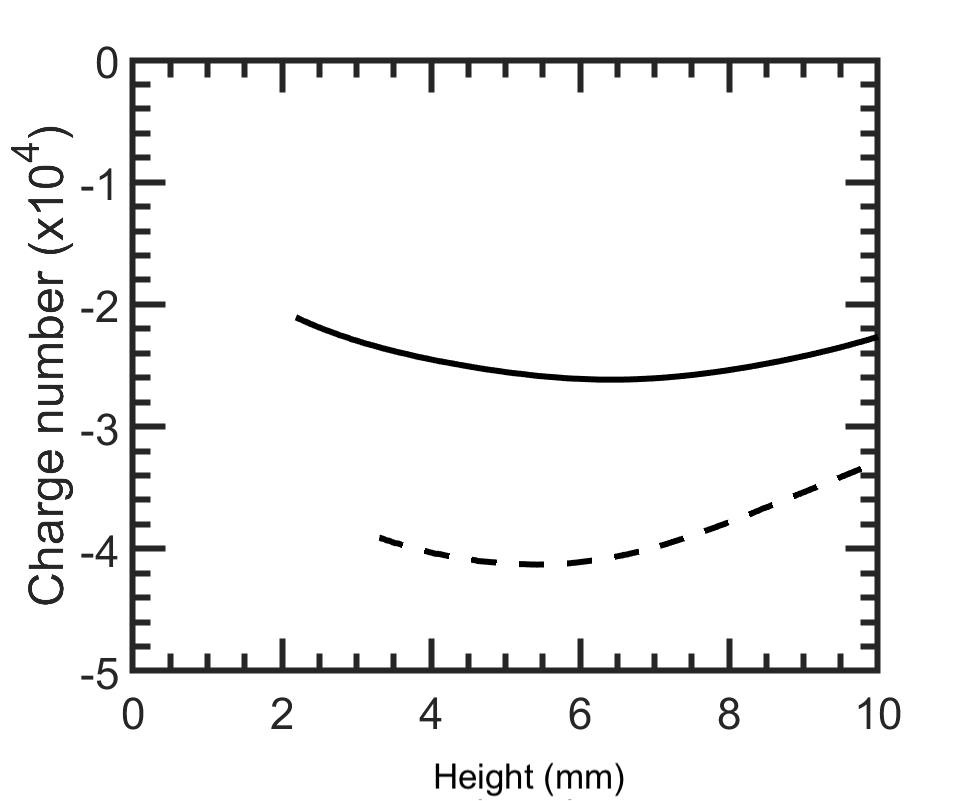}
    \caption{}
    \label{fig:47V_Q}
  \end{subfigure}
  ~
  \begin{subfigure}[t]{0.315\textwidth}
    \centering
    \includegraphics[scale=0.19]{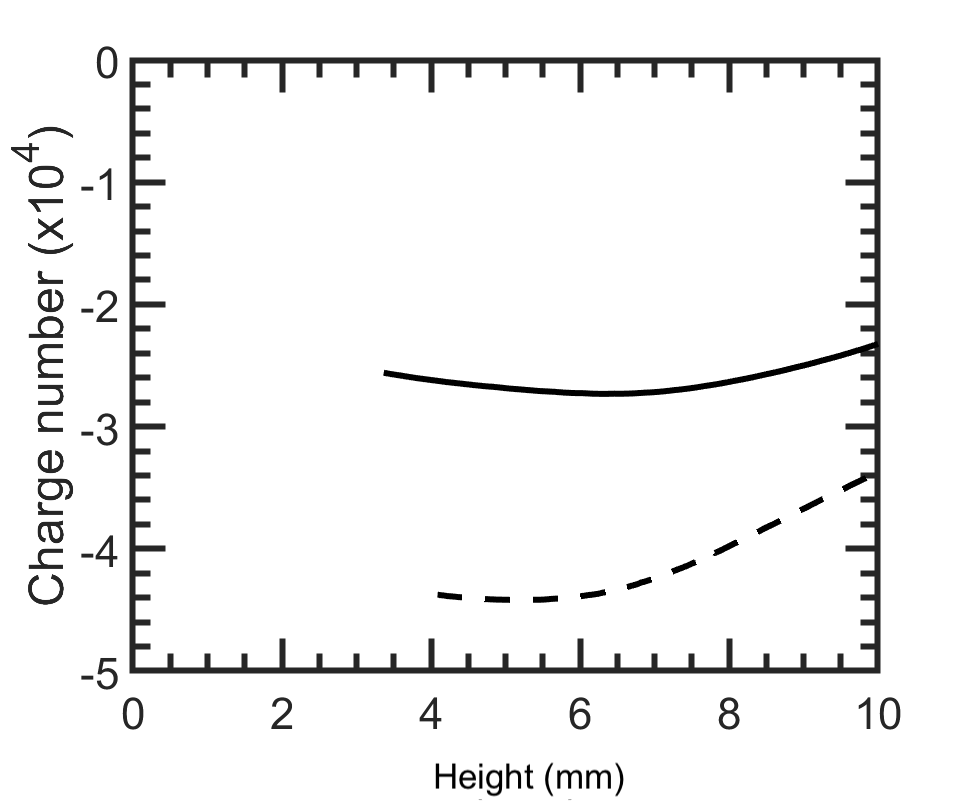}
    \caption{}
    \label{fig:64V_Q}
  \end{subfigure}
  \caption{Grain charge number on a 8.89 $\mu$m dust grain vs height above the electrode for the OML model (dashed lines) and CEC model (solid line); (a) 30 V plasma; (b) 47 V;  (c) 64 V.} 
  \label{fig:Dust_FeQ}
\end{figure}

\begin{figure}%[h]   
  \centering
  \begin{subfigure}[t]{0.315\textwidth}
    \centering
    \includegraphics[scale=0.19]{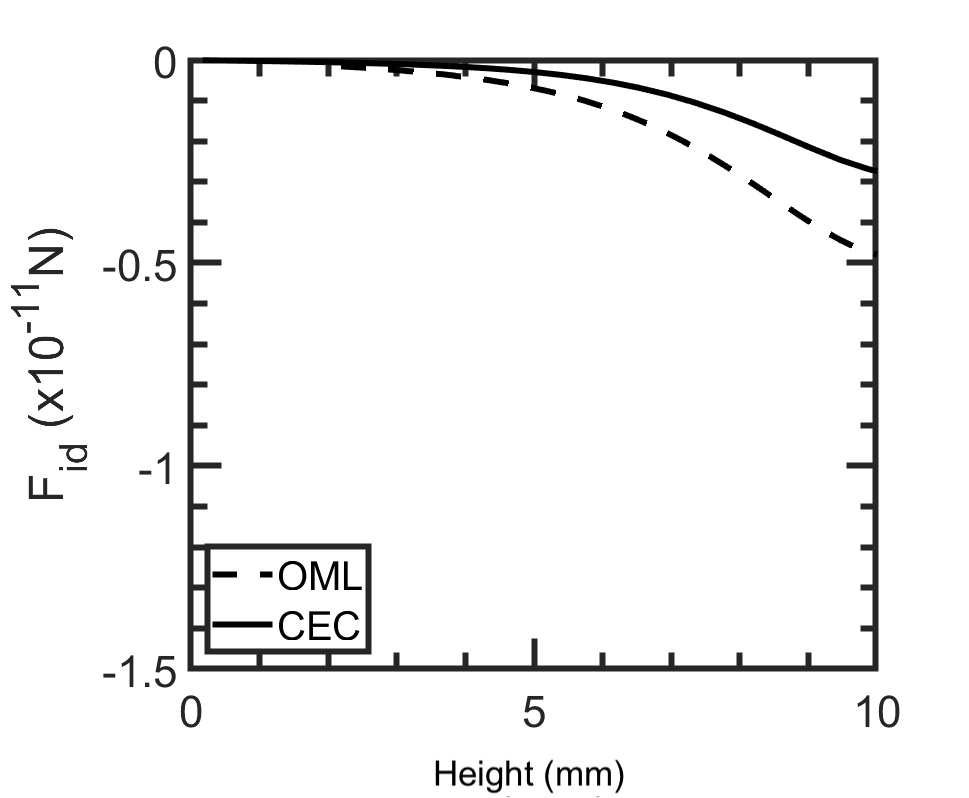}
    \caption{}
    \label{fig:30V_Fid}
  \end{subfigure}
  ~
  \begin{subfigure}[t]{0.315\textwidth}
    \centering
    \includegraphics[scale=0.19]{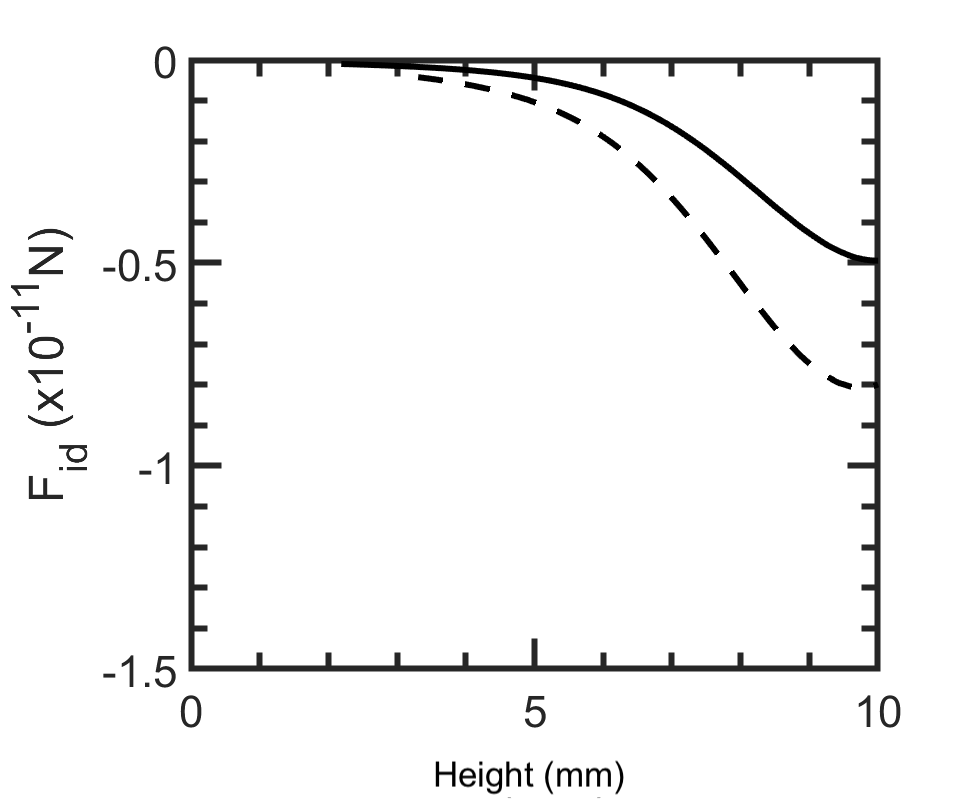}
    \caption{}
    \label{fig:47V_Fid}
  \end{subfigure}
  ~
  \begin{subfigure}[t]{0.315\textwidth}
    \centering
    \includegraphics[scale=0.19]{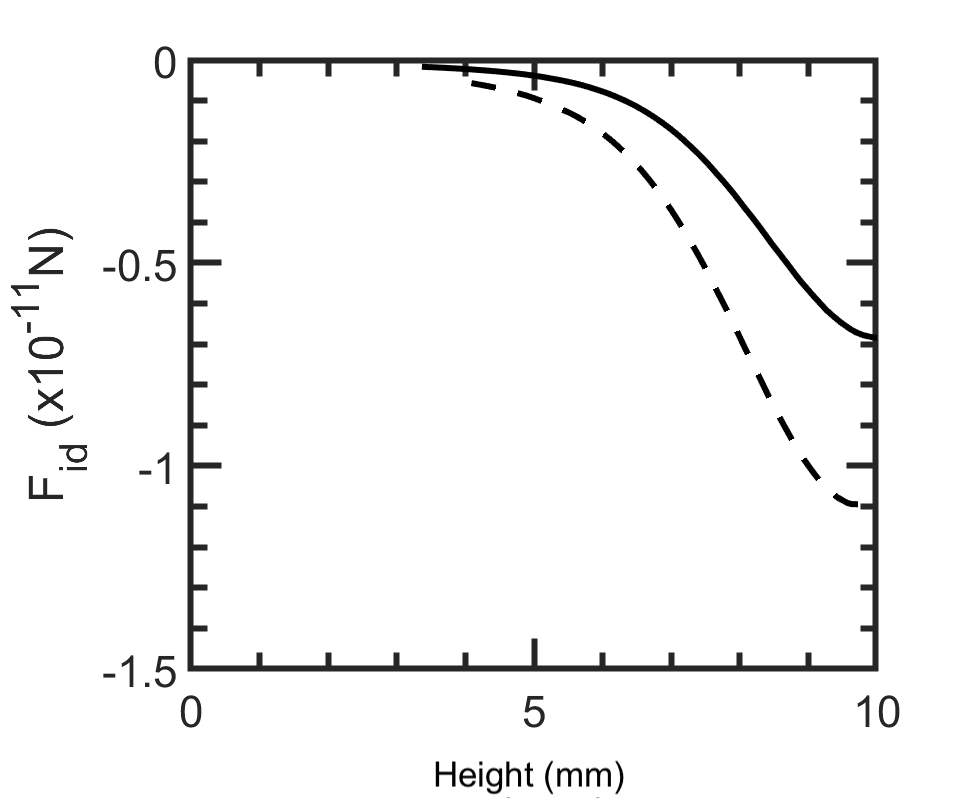}
    \caption{}
    \label{fig:64V_Fid}
  \end{subfigure}
  \caption{Ion drag force on a 8.89 $\mu$m dust grain vs height above the electrode for the OML model (dashed lines) and CEC model (solid line); (a) 30 V plasma; (b) 47 V;  (c) 64 V.} 
  \label{fig:Dust_FeFid}
\end{figure}

The difference in the electric force between the two models is due to a lower charge represented by the CEC model, as seen in Figure \ref{fig:Dust_FeQ}, which shows the dust charge number, i.e. the number of electrons collected on the dust grain $Q/{\lvert e \rvert}$. The additional ion current due to thermal ions in the CEC model reduces the dust charge by around 20\% throughout the domain. 
The ion drag force is also dependent on the grain charge, and Figure \ref{fig:Dust_FeFid} shows the ion drag force on the dust grain for the two charging models, OML and CEC. The ion drag force, Eq.~(\ref{eq:Fid}), is a function of the dust floating potential and the ion velocity, through the term $e(\phi_p-\phi)/{m_i v_s^2}$, so the difference between the OML and CEC results is larger in the regions near the bulk where the ion velocity is smaller and less dominant. Since the dust grains levitate between 4-8mm above the electrode, the ion drag force will be around 20-40\% of the electric force and will influence the levitation height of the dust grain.

\begin{figure}%[h]   
  \centering
  \begin{subfigure}[t]{0.315\textwidth}
    \centering
    \includegraphics[scale=0.19]{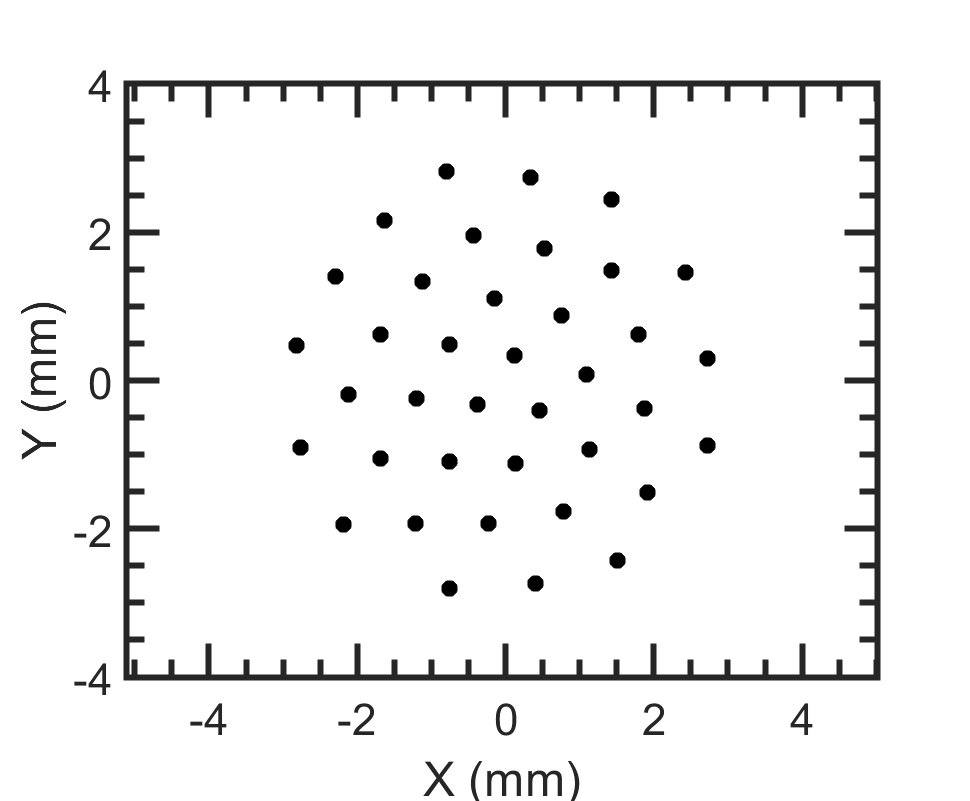}
    \caption{}
  \end{subfigure}
  ~
  \begin{subfigure}[t]{0.315\textwidth}
    \centering
    \includegraphics[scale=0.19]{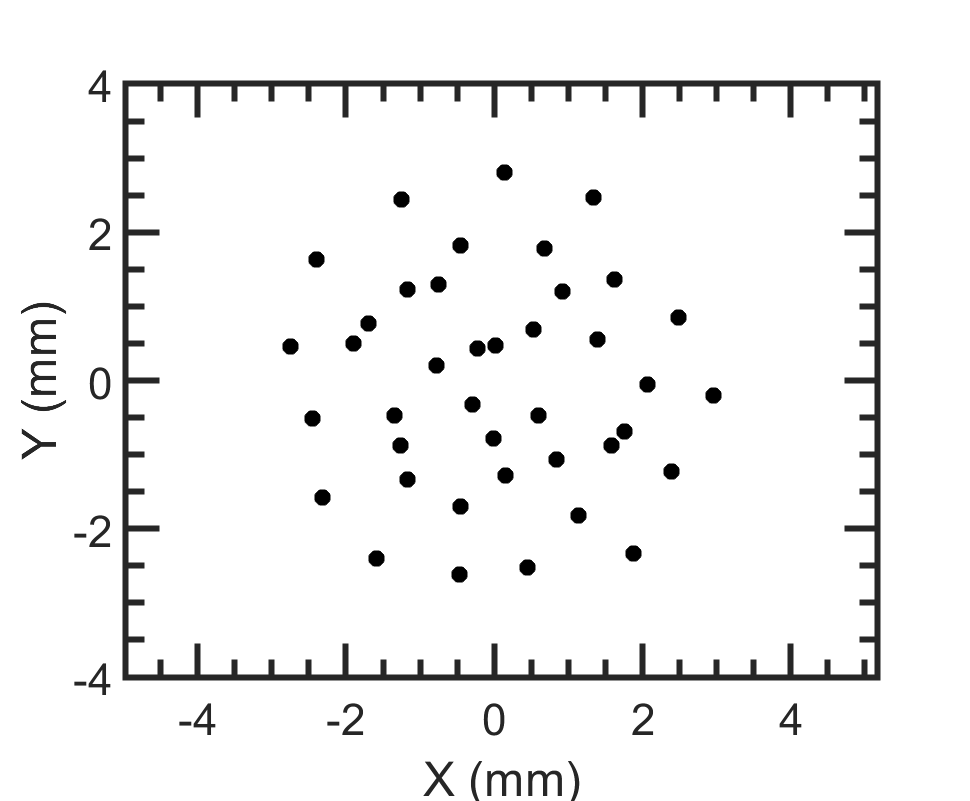}
    \caption{}
  \end{subfigure}
  
  \begin{subfigure}[t]{0.315\textwidth}
    \centering
    \includegraphics[scale=0.19]{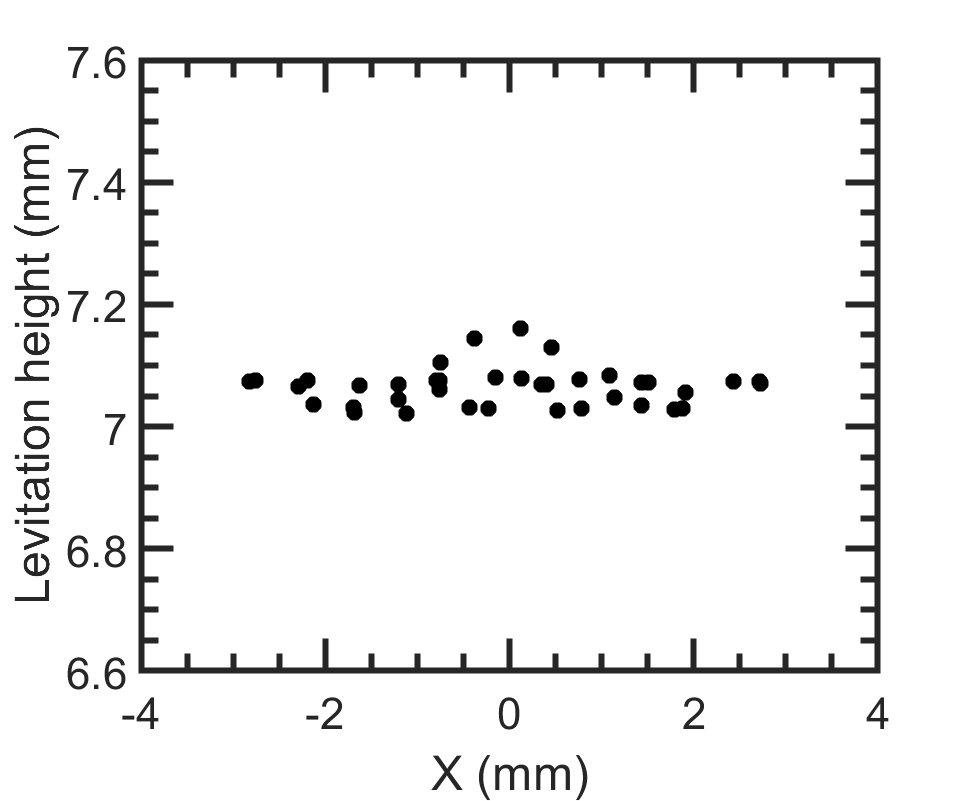}
    \caption{}
  \end{subfigure}
  ~
  \begin{subfigure}[t]{0.315\textwidth}
    \centering
    \includegraphics[scale=0.19]{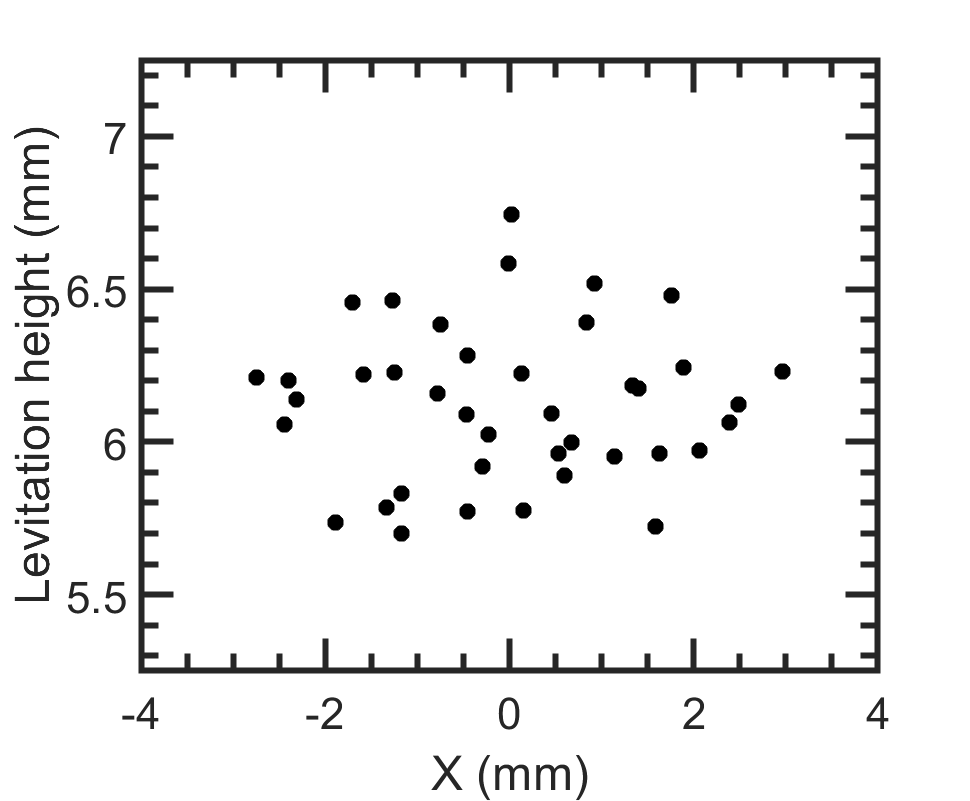}
    \caption{}
  \end{subfigure}
  \caption{Top and side views of dust cluster (a) top view of stable crystal; (b) top view of melted case; (c) side view of stable crystal; (d) side view of melted case. Both cases are 11.93 $\mu$m using OML, stable example is in 70V plasma, melted example is in 30V plasma.} 
  \label{fig:Top_view}
\end{figure}

%%\clearpage
%%\section{Dust crystal results}
Dust crystals were simulated by releasing 40-100 dust grains in the plasma, sometimes dust grains were pushed far enough from the center that they crossed the electrode cutout and were not incorporated into the crystal.
%%%%%%%%%%%% max crystal size
The resulting crystals consisted of a single layer. 
%The levitation height was stable for all cases. 
However, the in-plane arrangement of the dust grains was not always stable, some cases melted and did not form a crystalline structure.
Figure~\ref{fig:Top_view} shows top and side views of one of the stable crystals as well as one of the melted cases. 
The stable crystal has a hexagonal structure with each grain having six neighbors, as is typical for a strongly coupled two-dimensional Yukawa system~\cite{Klumov2010}, the inter-particle distance does not vary substantially across the crystal. The levitation height varies by 0.13 mm across the crystal due to the radial variation in the plasma resulting from the electrode cutout. 
The levitation height for the grains in the melted case varies by 1.0 mm and the grains move throughout the crystal.
Table~\ref{tab:melting} gives the breakdown of which cases melted. The CEC model was more susceptible to melting than the OML model. For both models, melting was related to both grain size and the RF voltage applied to the plasma, with larger grains and lower voltages being more likely to melt.
Since dust grains in the melted cases had a height variation on the order of 1 mm and were in motion,  accurately representing the melted cases requires for the 3D dynamics to be treated with a more advance wake model.  Therefore, the melted cases are not discussed further considering that this work is focused on the stable crystals.

\begin{figure}[h]   
  \centering
  \begin{subfigure}[t]{0.315\textwidth}
    \centering
    \includegraphics[scale=0.19]{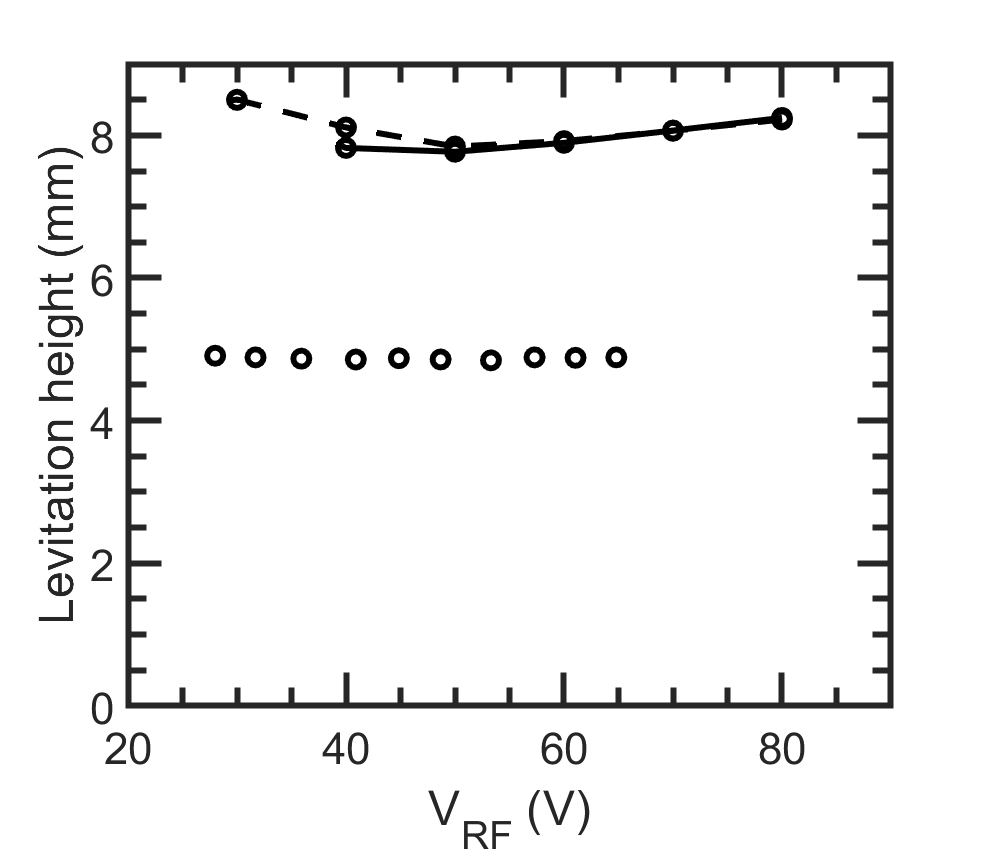}
    \caption{}
    \label{fig:637_Z}
  \end{subfigure}
  ~
  \begin{subfigure}[t]{0.315\textwidth}
    \centering
    \includegraphics[scale=0.19]{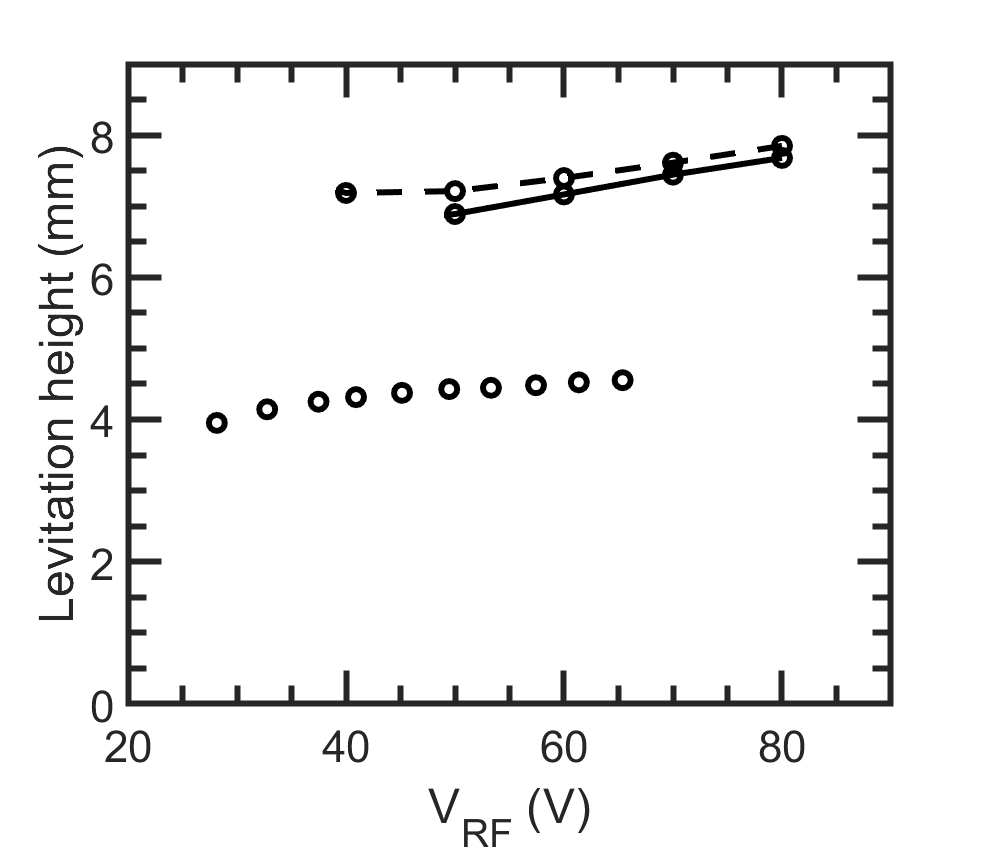}
    \caption{}
    \label{fig:889_Z}
  \end{subfigure}
  ~
  \begin{subfigure}[t]{0.315\textwidth}
    \centering
    \includegraphics[scale=0.19]{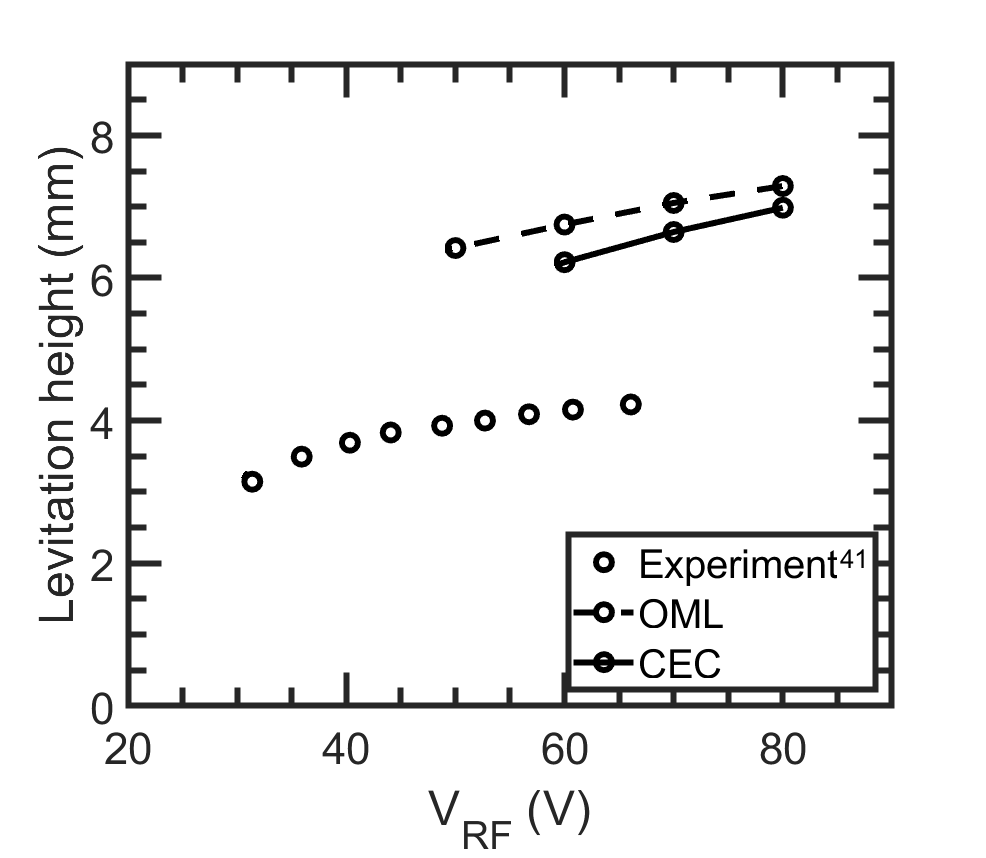}
    \caption{}
    \label{fig:1193_Z}
  \end{subfigure}
  \caption{Levitation heights vesrus RF voltage in experiments of \citet{Douglass2012} and the OML  and CEC models in the current study for a grain diameter of (a) 6.37 $\mu m$; (b) 8.89 $\mu m$; (c) 11.93 $\mu m$. } 
  \label{fig:Dust_Z}
\end{figure}

The levitation height of the dust grains above the electrode is shown in Figure \ref{fig:Dust_Z}. 
The effect of the collision enhanced charging is not substantial with the CEC model producing 0.5 mm lower levitation heights compared to the OML model for the largest grains and having negligible effect on the levitation heights of the smallest grains.
Both the OML and CEC models over-predicted the levitation height by around 2mm compared to the experimental results which is consistent with the visual inspection of Figure~\ref{fig:Dust_Fe} which indicated the models had a 1-2mm wider sheath than the experiment. 
%The difference in height between the OML and CEC models is larger for small voltages and large dust grains, but the overall effect is small for most cases. 

\begin{figure}[h]   
  \centering
  \begin{subfigure}[t]{0.315\textwidth}
    \centering
    \includegraphics[scale=0.19]{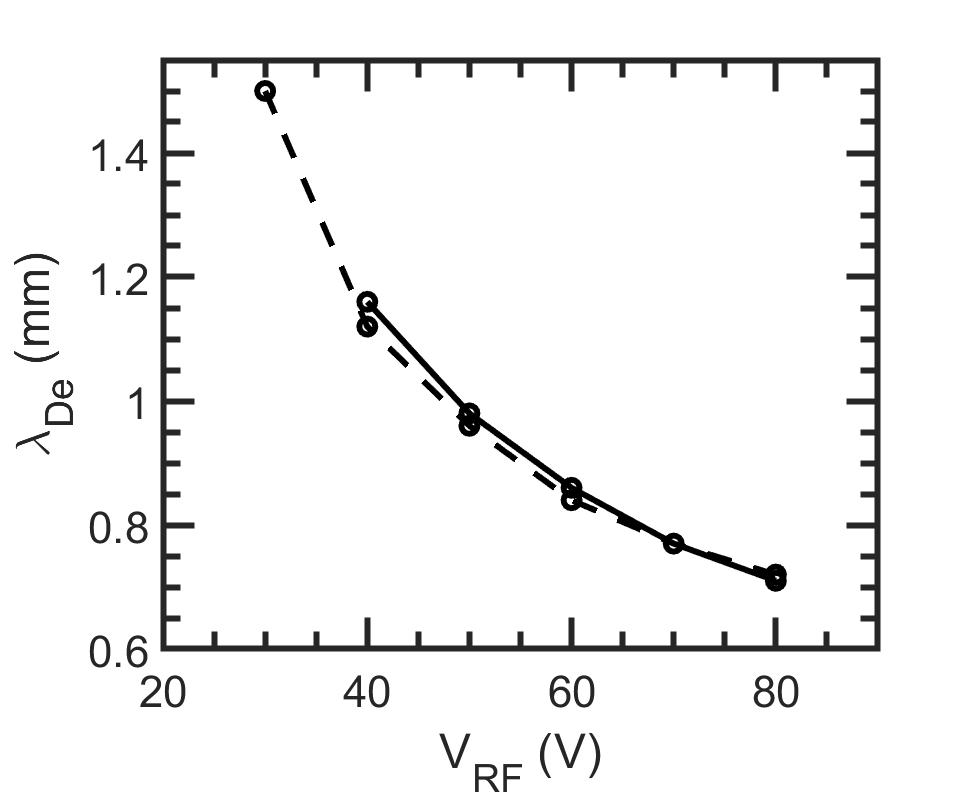}
    \caption{}
    \label{fig:637_debye}
  \end{subfigure}
  ~
  \begin{subfigure}[t]{0.315\textwidth}
    \centering
    \includegraphics[scale=0.19]{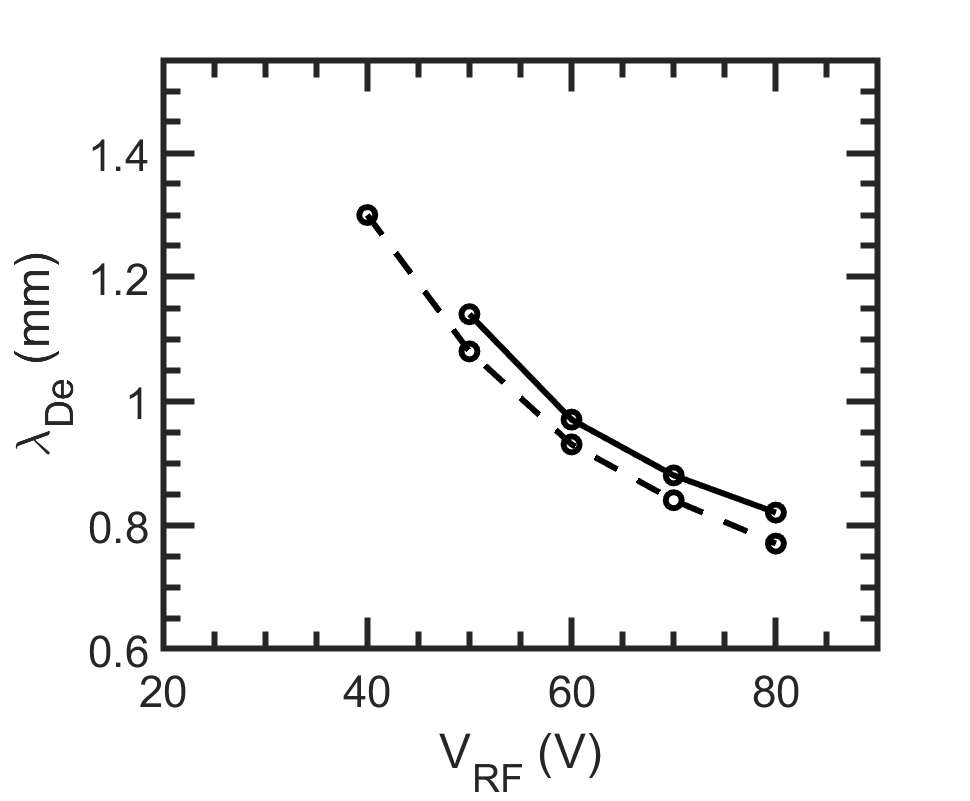}
    \caption{}
    \label{fig:889_debye}
  \end{subfigure}
  ~
  \begin{subfigure}[t]{0.315\textwidth}
    \centering
    \includegraphics[scale=0.19]{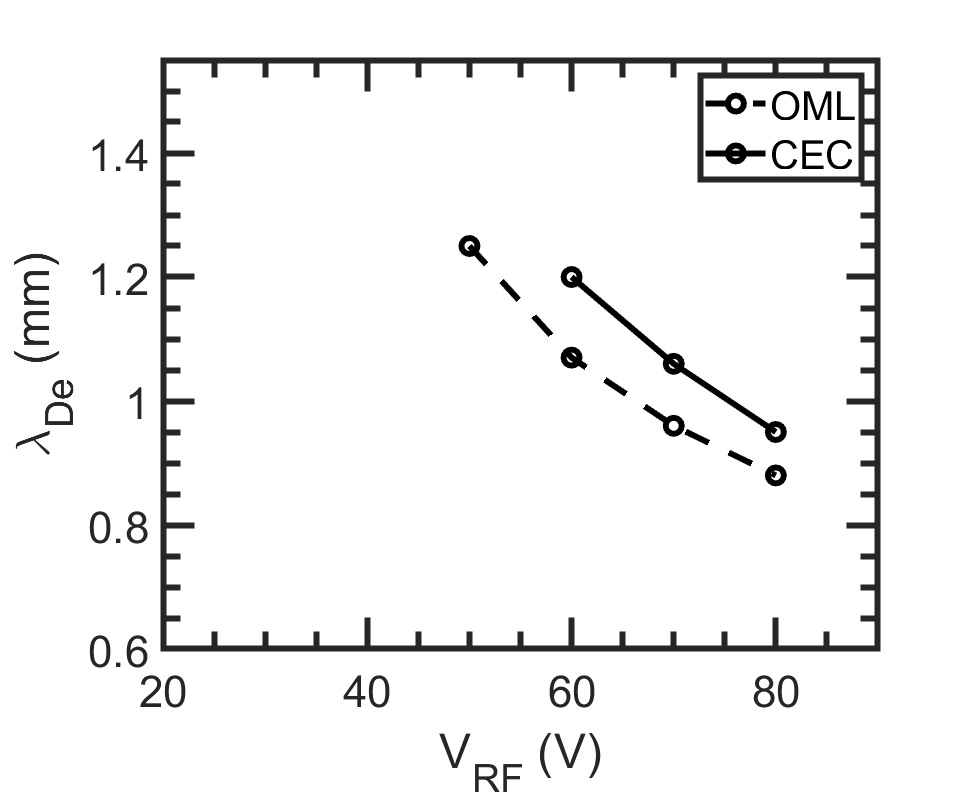}
    \caption{}
    \label{fig:1193_debye}
  \end{subfigure}
  \caption{Debye length versus RF voltage at the crystal location for OML model (dashed line) and CEC model (solid line)   for a grain diameter of (a) 6.37 $\mu m$; (b) 8.89 $\mu m$; (c) 11.93 $\mu m$. } 
  \label{fig:Dust_debye}
\end{figure}

The electron Debye length at the location of the crystals is shown in Figure~\ref{fig:Dust_debye}. The CEC model has a higher Debye length because the grains have a lower levitation height. The Debye length is larger than the variation in the levitation height across a stable crystal, by a factor of 5, so the chosen wake model is sufficient to handle the stable crystals in spite of the wake model's limitations in handling 3D dynamics.

\begin{figure}%[h]   
  \centering
  \begin{subfigure}[t]{0.315\textwidth}
    \centering
    \includegraphics[scale=0.19]{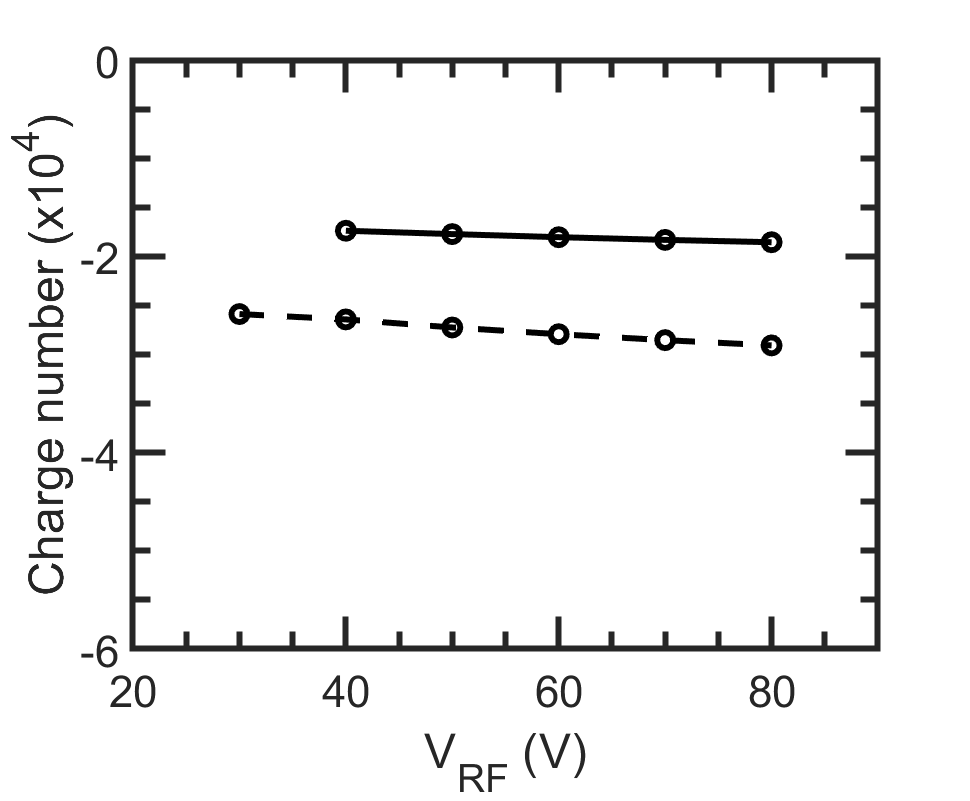}
    \caption{}
    \label{fig:637_Q}
  \end{subfigure}
  ~
  \begin{subfigure}[t]{0.315\textwidth}
    \centering
    \includegraphics[scale=0.19]{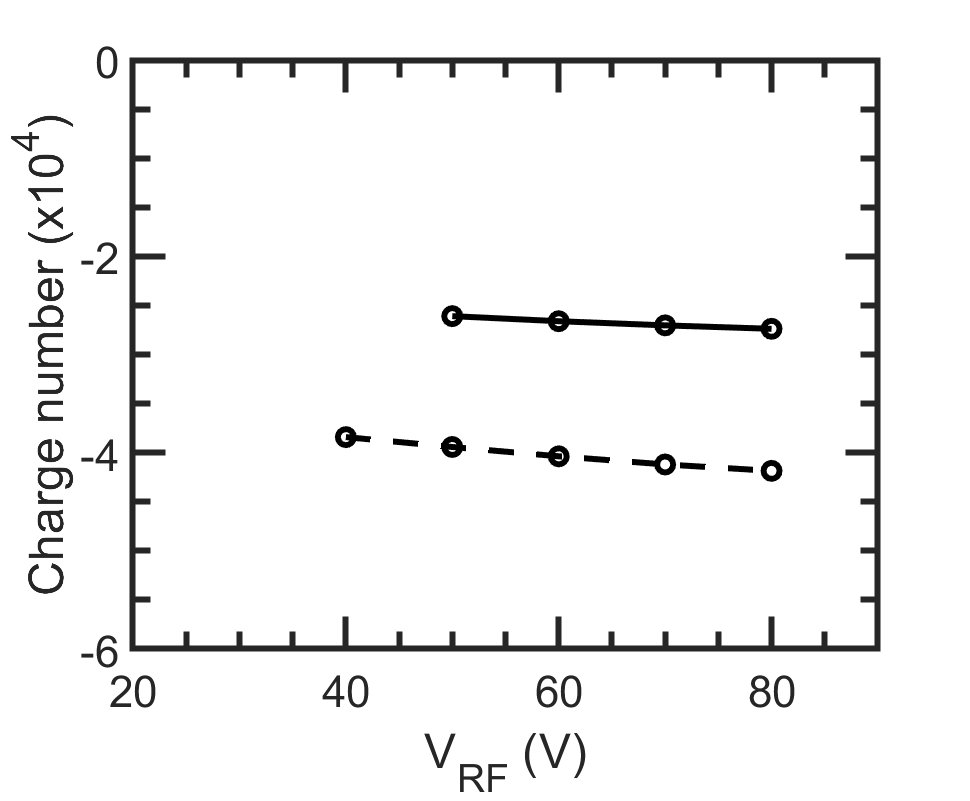}
    \caption{}
    \label{fig:889_Q}
  \end{subfigure}
  ~
  \begin{subfigure}[t]{0.315\textwidth}
    \centering
    \includegraphics[scale=0.19]{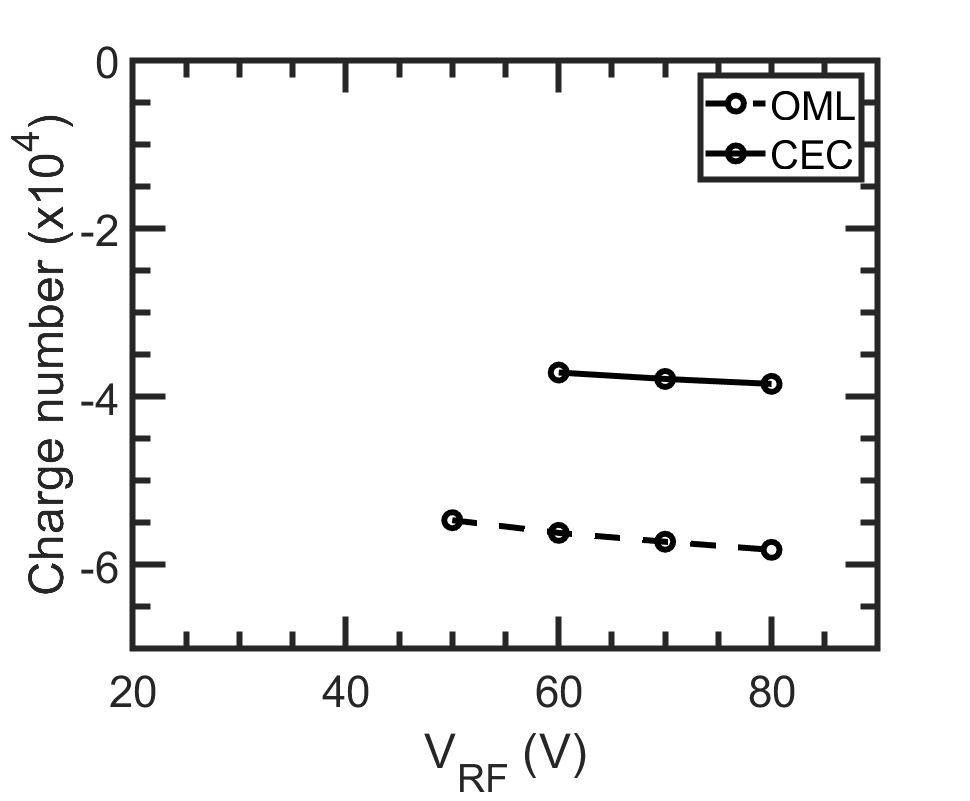}
    \caption{}
    \label{fig:1193_Q}
  \end{subfigure}
  \caption{Grain charge number using OML (dashed line) and CEC (solid line) charging models as a function of RF voltage for a grain diameter of (a) 6.37 $\mu m$; (b) 8.89 $\mu m$; (c) 11.93 $\mu m$.} 
  \label{fig:Dust_charge}
\end{figure}

The average charge number of the dust grains is shown in Figure \ref{fig:Dust_charge}. The charge increases as the dust grain radius is increased but is less sensitive to the applied RF voltage. As expected, the OML model produces a higher charge, around 20-30\% higher than the CEC model. 
The lower charge produced by the CEC model will reduce the electric force as well as the ion drag force.
The large difference in the charge and small difference in the levitation height indicates that the reduction in the electric force is balanced by the reduction in the ion drag force.

\begin{figure}%[h] 
  \centering
  \begin{subfigure}[t]{0.315\textwidth}
    \centering
    \includegraphics[scale=0.19]{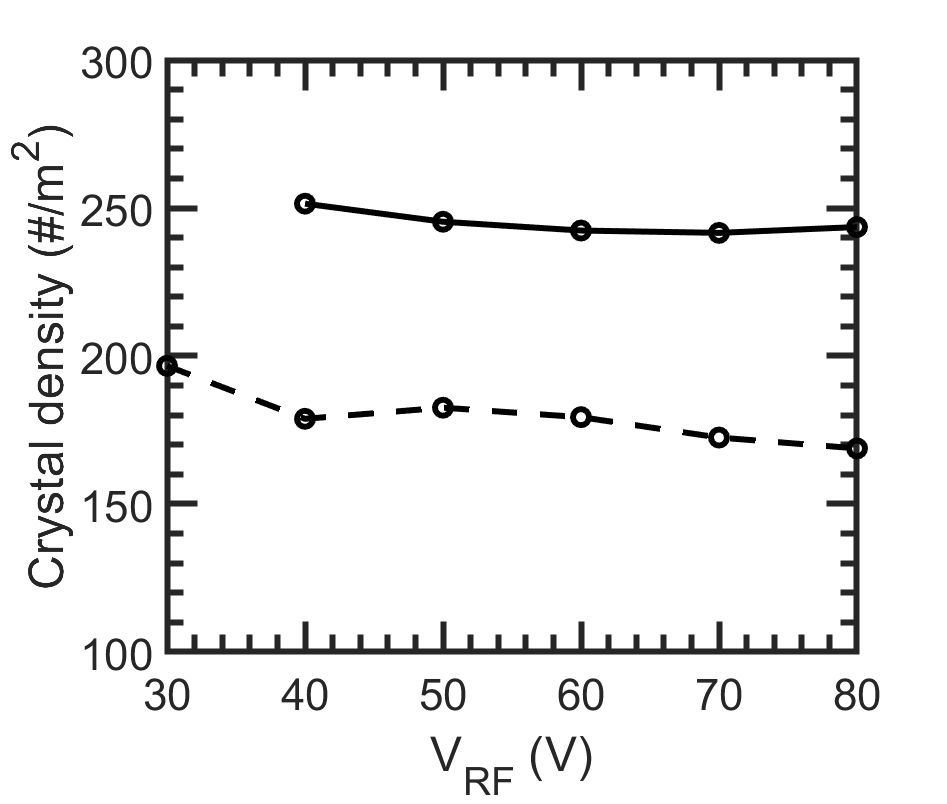}
    \caption{}
    \label{fig:637_density}
  \end{subfigure}
  ~
  \begin{subfigure}[t]{0.315\textwidth}
    \centering
    \includegraphics[scale=0.19]{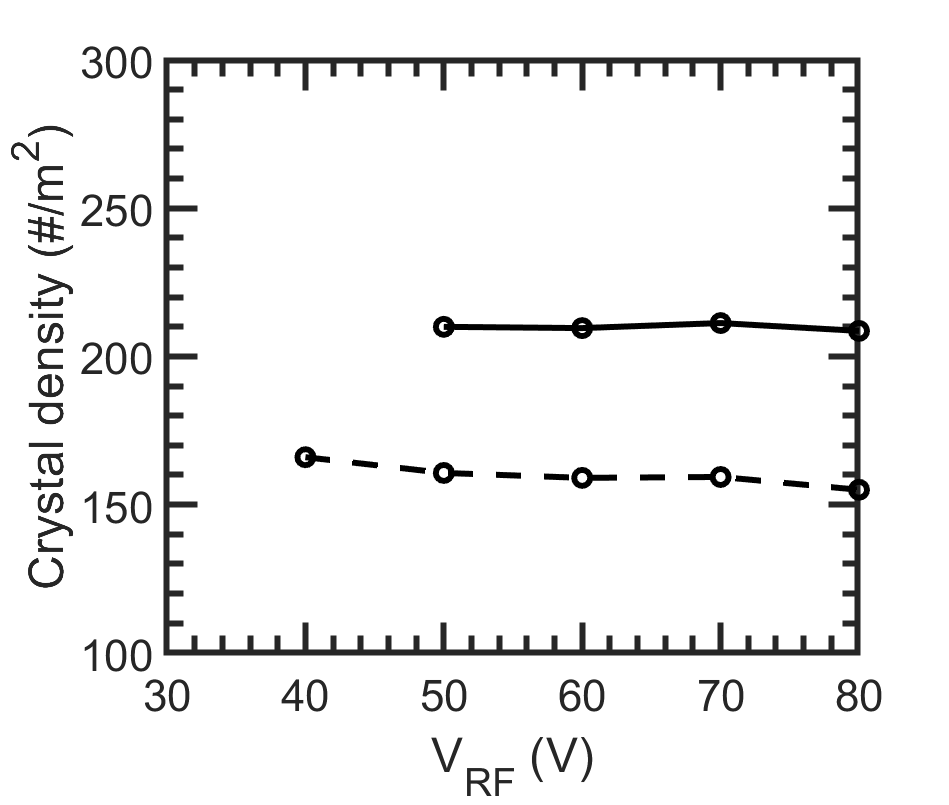}
    \caption{}
    \label{fig:889_density}
  \end{subfigure}
  ~
  \begin{subfigure}[t]{0.315\textwidth}
    \centering
    \includegraphics[scale=0.19]{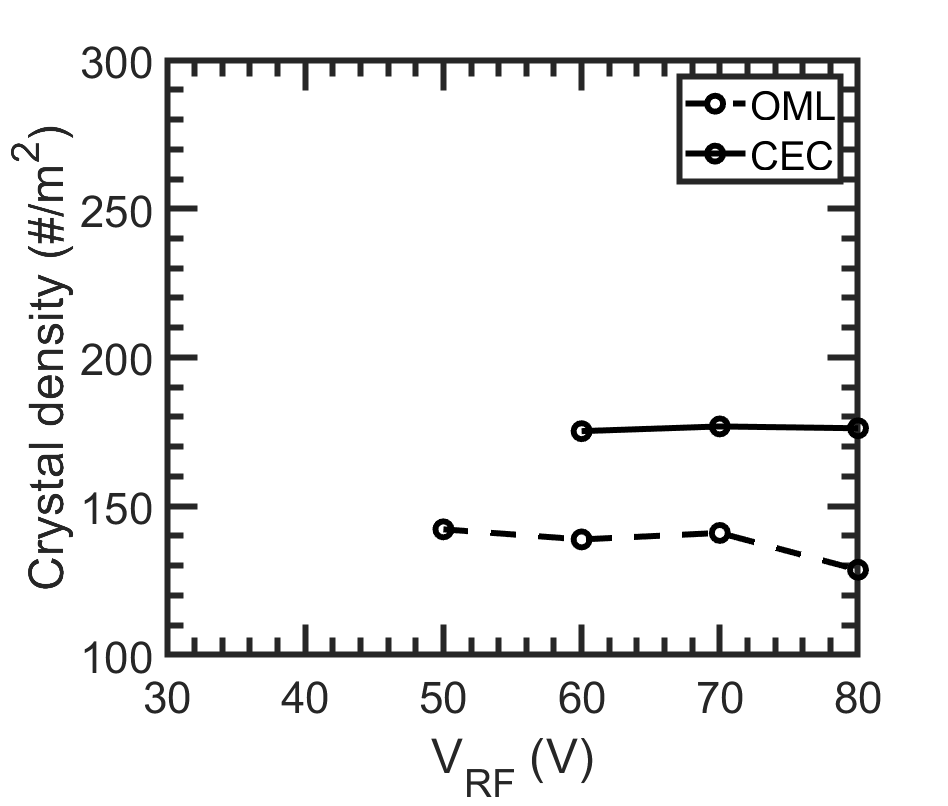}
    \caption{}
    \label{fig:1193_density}
  \end{subfigure}
  \caption{Crystal density using OML (dashed line) and CEC (solid line) charging models as a function of RF voltage for a grain diameter of (a) 6.37 $\mu m$; (b) 8.89 $\mu m$; (c) 11.93 $\mu m$.} 
  \label{fig:Crystal_density}
\end{figure}

Since the crystals were single layer, the crystal density was defined as the number of dust grains divided by the area of the circle enclosing the crystal. 
%Cases that melted were excluded and 
The density of the stable crystals is shown in Figure~\ref{fig:Crystal_density}. The density decreased for larger dust grains but did not show a strong dependence on the applied RF voltage due to the fact that the charge was not highly sensitive to the voltage. The CEC model produced a 20-30\% higher density than the OML model since the lower charge in the CEC model results in weaker repulsion between the grains.

\begin{figure}%[h] 
  \centering
  \begin{subfigure}[t]{0.315\textwidth}
    \centering
    \includegraphics[scale=0.19]{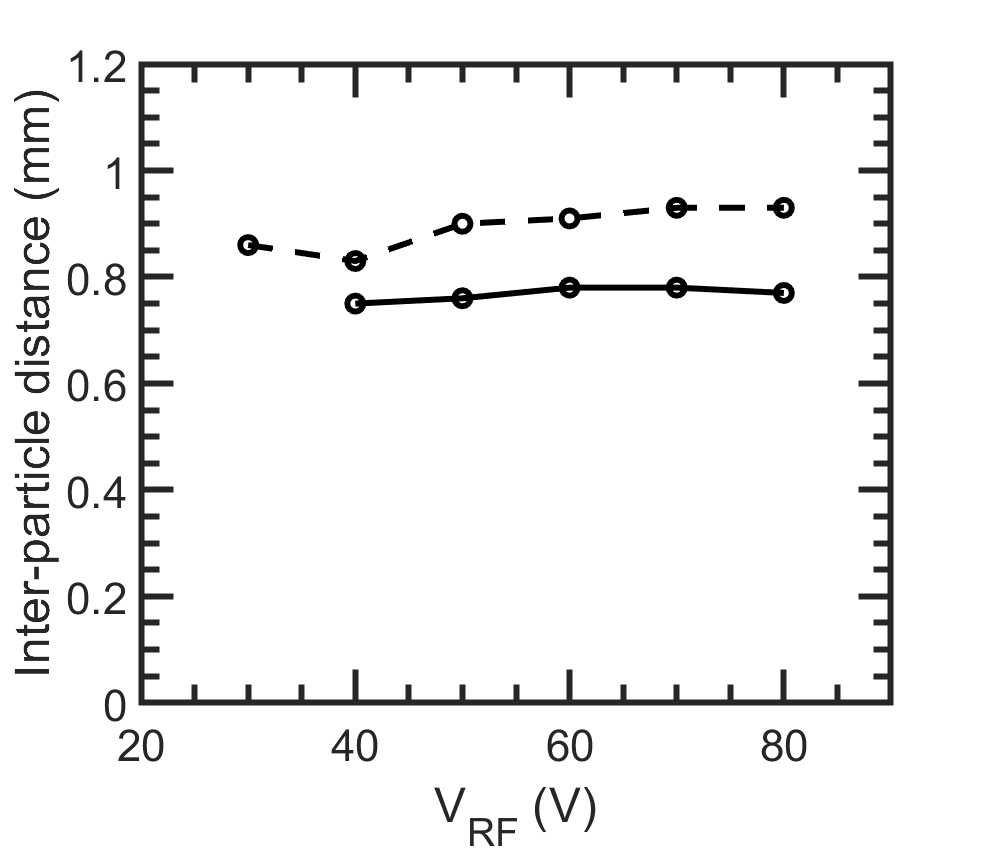}
    \caption{}
    \label{fig:637_dist}
  \end{subfigure}
  ~
  \begin{subfigure}[t]{0.315\textwidth}
    \centering
    \includegraphics[scale=0.19]{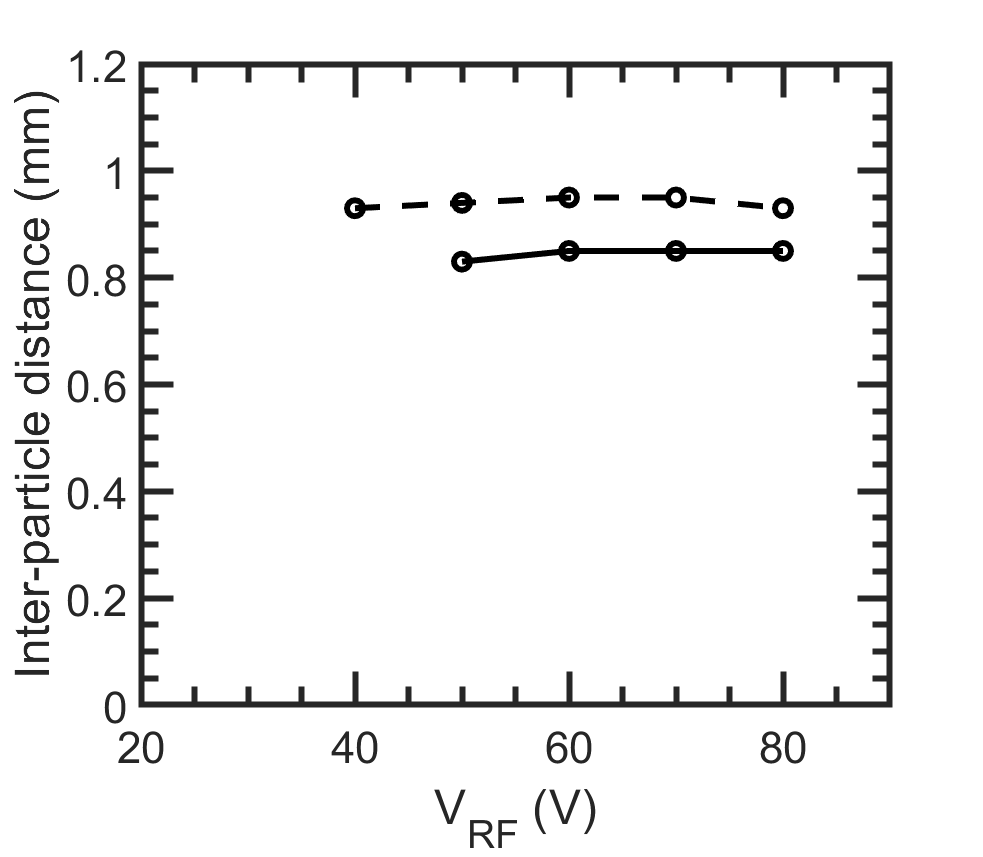}
    \caption{}
    \label{fig:889_spacing}
  \end{subfigure}
  ~
  \begin{subfigure}[t]{0.315\textwidth}
    \centering
    \includegraphics[scale=0.19]{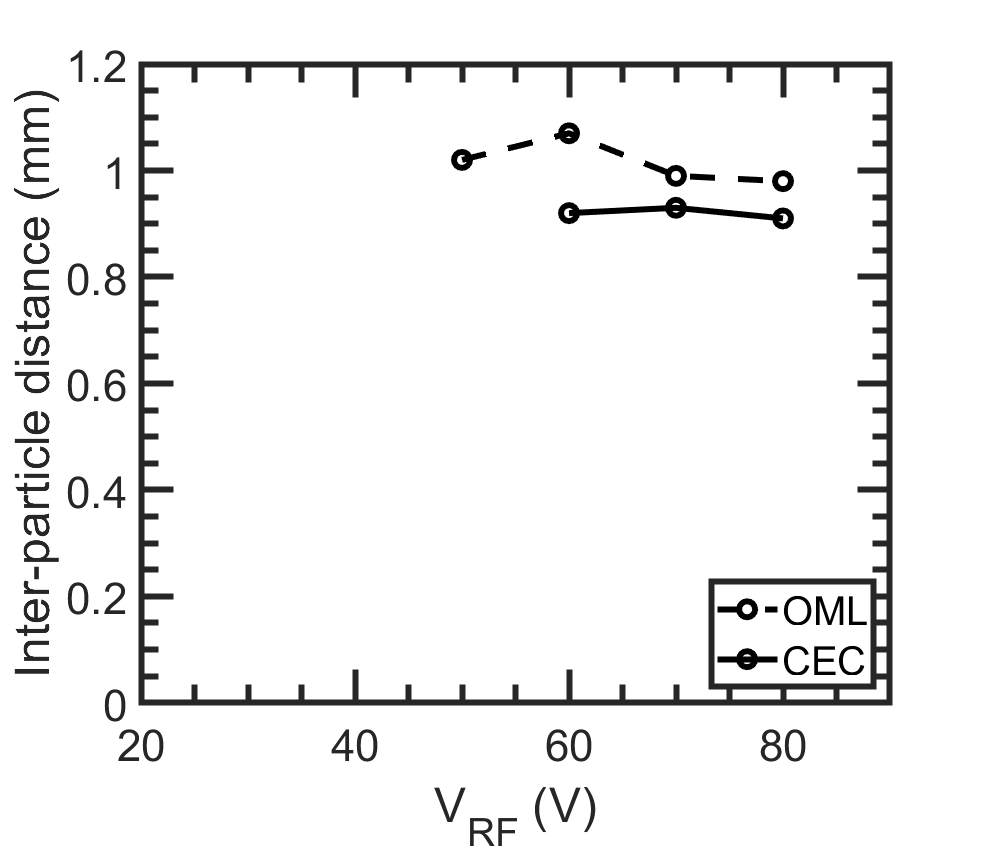}
    \caption{}
    \label{fig:1193_spacing}
  \end{subfigure}
  \caption{Inter-particle spacing using OML (dashed line) and CEC (solid line) charging models as a function of RF voltage for a grain diameter of (a) 6.37 $\mu m$; (b) 8.89 $\mu m$; (c) 11.93 $\mu m$.} 
  \label{fig:Crystal_spacing}
\end{figure}

Figure~\ref{fig:Crystal_spacing} shows the inter-particle distance which is the distance between a grain and its neighbors.%, excluding cases that melted. 
It is determined by creating a histogram of the distance between each pair of grains, the first peak occurs at the inter-particle distance. The inter-particle distance does not show a strong dependence on either the applied RF voltage or the dust grain size. Including the collision enhancement to the ion current reduced the inter-particle distance by 10-15\%.

\section{Summary and Conclusions}  %%% Conclusions
\label{sec:Conclusions}
Monolayer dust crystals  were investigated by modeling in a computational configuration resembling the GEC cell of previous experiments with dust particles \cite{Douglass2012}. The plasma was modeled using a continuum description and the back way coupling effect of dust on the plasma was neglected due to the low dust loading. Once the plasma had reached quasi-steady state, the plasma values were averaged over one RF cycle and the dust grains were added. The dust was released into the plasma and tracked over time. The time history of the individual grains was examined and the final crystal structure was investigated. Two models for determining the dust grain charge, viz. the OML and CEC models, were implemented and investigated against each other and the data from \citet{Douglass2012}. Since the CEC charging model is simplified to the OML charging model at the collisionless limit,  the comparison of the results obtained from two models shedded a light on the influence of the collision-enhanced ion current in  dust crystals.

Comparing trajectories for a single grain showed that both models had a less pronounced transition point between the bulk and sheath and the transition occurred 1-2mm higher than in the experimental data. The CEC model had a lower dust charge than the OML model due to the increased ion current. The lower charge in the CEC model resulted in lower values for both the electric and ion drag forces. Consequently, the CEC results agreed better with the experimental electric force profile. 

An examination of clusters showed that larger grains in low voltage plasmas did not always form crystalline structures, and the CEC model was more likely not to form a crystal.
The stable crystals were examined further but the melted cases were not considered further due to the model limitations in dealing with melted crystals.
Stable crystals were two-dimensional and had a hexagonal structure, the side view of the crystal showed a small amount of curvature due to the parabolic confining potential created by the electrode cutout. 
The variation in the levitation height was around 0.1-0.15mm which is substantially smaller than the Debye length, which was on the order of 1mm, so the structure was essentially 2D. The difference between the modeled and previously measured levitation heights remained below 4mm. 

%However, the levitation height and dust grain charge were stable regardless of whether a crystal was formed. 
While the CEC model produced 20-30\% lower charges, the reduction in the ion drag force balanced the reduction in the electric force for most cases so the levitation height was similar for both the CEC and OML models. 
Both OML and CEC models overpredicted the levitation height compared to the experimental data, which could be a result of the wider sheath observed in the single grain validation study.
The levitation height decreased as the voltage was reduced or the dust grain radius was increased.
%An examination of the horizontal structure of the dust crystals 
The crystal density decreased as the dust radius increased but did not show a strong dependence on the RF voltage. The inter-particle distance increased by around 0.1 mm as the dust radius increased.
The CEC model produced a 10-15\% smaller inter-particle distance and 20-30\% larger crystal density than the OML model. Both  interparticle distance and crystal density were negligibly sensitive to the applied voltage. 

\section*{Data Availability Statement}
The data that support the findings of this study are available from the corresponding author upon reasonable request.

\appendix %%%%%%%%%%%%%% move potential and derivation to here, add numerical discussion
{\section{Trapping region details}
\label{sec:R*Z*Derive}}
The CEC model accounts for the increased ion current due to charge exchange collisions creating thermal ions. If the potential energy of the ion is below a certain threshold, it will be collected by the dust grain. The potential energy of the ion is defined in a polar coordinate system centered on the dust grain with $\theta=0$ in the direction of the ion drift. The ion potential energy is given by
\begin{equation} \label{eq:phi_ion}
e\Phi(r,\theta)=\frac{e q_d}{4\pi \epsilon_0 r}-e E_\mathrm{z}r\cos(\theta)
\end{equation}
where $q_d$ is the dust charge and $E_z$ is the electric field. 
The trapping criteria is:
\begin{equation} \label{eq:criteria}
e\Phi(r,\theta)+k_B T_i<e\Phi(z_1,0)
\end{equation}
where $z_1$ is the saddle point. \citet{Piel2015} approximated the saddle point by setting the electric force equal to the gravitational force which assumes the dust grain has already settled to the correct location. For this study, the mathematical saddle point is used instead since the dust grain is moving.

By applying the definition of the ion potential energy, Eq.~(\ref{eq:phi_ion}), the trapping criteria can be expanded as:
\begin{equation} \label{eq:criteria2}
\frac{e q_d}{4\pi \epsilon_0 r}-e E_\mathrm{z}r\cos(\theta)+k_B T_i=\frac{e q_d}{4\pi \epsilon_0 z_1}-e E_\mathrm{z}z_1 .
\end{equation}
This can be simplified by defining $A=\frac{e q_d}{4\pi \epsilon_0}$, $B=e E_\mathrm{z}$, and $C=\frac{e q_d}{4\pi \epsilon_0 z_1}-e E_\mathrm{z}z_1-k_B T_i$, which allows the trapping condition to be written as:
\begin{equation}
\frac{A}{r}-Br\cos(\theta)=C .
\end{equation}

The trapping region is ellipsoidal and described by the major semi-axis, $Z^*$, and minor semi-axis, $R^*$. Both terms, $Z^*$ and $R^*$, are determined numerically.
The trapping region is not centered on the dust grain but is centered downstream of the dust grain due to the ion wake.
The major semi-axis occurs along the x-axis, while the minor semi-axis is parallel to, but not along, the $y$-axis.

The major semi-axis, $Z^*$, is half the distance between the maximum and minimum $x$-values, $Z^*=0.5*(x_{max}-x_{min})$, which are determined from:
\begin{equation} \label{eq:xmax}
\frac{A}{x_{max}}-B x_{max}\cos(0^\circ)=C
\end{equation}
and
\begin{equation} \label{eq:xmin}
\frac{A}{x_{min}}-Bx_{min}\cos(180^\circ)=C .
\end{equation}

The minor semi-axis,$R^*$, is half the distance between the maximum and minimum y-values, which means it is necessary to solve $\frac{A}{r}-Br\cos(\theta)=C$ while maximizing $y=\pm r\sin\theta$.
By using Lagrangian multipliers, the angle for $y_{max}$ is related to $r_y$ by:
\begin{equation} \label{eq:theta_ymax}
\cos(\theta_y)=\frac{-B}{A}r_y^2 ,
\end{equation}
and the value for $r_y$ can be obtained from:
\begin{equation} \label{eq:r_ymax}
\frac{A}{r_y}+\frac{B^2}{A}r_y^3=C .
\end{equation}
Since the trapping region is symmetric about the x-axis, $y_{min}=-y_{max}$ and $R^*=r_y\sin\theta_y$.

%\section*{References}

\bibliography{Dust_references}

\newpage
%%%% Tables and Figures

\begin{table}
  \centering
  \caption{Plasma parameters}
  \label{tab:parameters}
  \begin{tabular}{|l|c|}
   \hline
   Parameter & Value \\
   \hline
  Applied RF voltage, $V_\text{RF}$ & 30-80 V \\
  DC bias voltage, $V_\text{DC}$ & -5 V \\
  Frequency, $f$ & 13.56 MHz \\
  Neutral gas pressure, $P_\text{gas}$ &  20 Pa \\
  Ion and neutral temperature, $T_i=T_\text{gas}$ & 300 K \\
  Electron mobility $\mu_e P_\text{gas}$ & 4$\times$10$^3$ $\frac{\mathrm{m}^2}{\mathrm{V}\,\mathrm{s}}\mathrm{Pa}$\\
  Electron diffusion $D_e P_\text{gas}$ & 1.6$\times$10$^4$ $\frac{\mathrm{m}^2}{\mathrm{s}}\mathrm{Pa}$\\
  Ion mobility $\mu_i P_\text{gas}$ & 1.86$\times$10$^1$ $\frac{\mathrm{m}^2}{\mathrm{V}\,\mathrm{s}}\mathrm{Pa}$ \\
  Ion diffusion $D_i P_\text{gas}$ & 5.33$\times$10$^{-1}$ $\frac{\mathrm{m}^2}{\mathrm{s}}\mathrm{Pa}$ \\
   \hline
  \end{tabular}
\end{table}
\begin{table}
  \centering
  \caption{Cases with Melting}
  \label{tab:melting}
  \begin{tabular}{|c|c|c|c|c|c|c|}
   \hline
    {} & \multicolumn{3}{c|}{OML} & \multicolumn{3}{c|}{CEC} \\
    \cline{2-7}
    {} & 6.37$\mu m$ & 8.89$\mu m$ & 11.93$\mu m$ & 6.37$\mu m$ & 8.89$\mu m$ & 11.93$\mu m$ \\
   \hline
  30V & & Melted & Melted & Melted & Melted & Melted \\
  40V & & & Melted & & Melted & Melted \\
  50V & & & & & & Melted \\
  60V & & & & & &  \\
  70V & & & & & &  \\
  80V & & & & & &  \\
   \hline
  \end{tabular}
\end{table}

\newpage

\clearpage

\clearpage

% Dust figures - single grain
%
\clearpage
\clearpage
\clearpage

% Crystal figures
\clearpage
\clearpage  
\clearpage
\clearpage

\clearpage
\clearpage

\end{document}